\documentclass[
 reprint,
 prx,
 amsmath,amssymb,
 aps,
 longbibliography,
floatfix
]{revtex4-1}
\usepackage[utf8]{inputenc}
\usepackage{lipsum}
\usepackage{amsmath}
\usepackage{mathbbol}
\usepackage{amssymb}             
\usepackage{amsthm}
\usepackage{xcolor}
\usepackage{ulem} %
\usepackage{siunitx}
\usepackage{graphicx}%
\usepackage{dcolumn}%
\usepackage{bm}%
\usepackage{physics}
\usepackage{hyperref}
\definecolor{bleuf}{rgb}{0,0.44,0.72}
\hypersetup{
    colorlinks=true,                          
    linkcolor=bleuf, %
    citecolor=bleuf, %
    urlcolor=bleuf  } 
\DeclareSymbolFontAlphabet{\amsmathbb}{AMSb}%

\newcommand{\beq}{\begin{equation}}
\newcommand{\eeq}{\end{equation}}

\newcommand{\BZ}{{\text{BZ}}}

\newcommand{\david}[2]{{ #2 }}

\begin{document}

\title{  
Geometry and  Topology Tango in Ordered and Amorphous Chiral Matter}

\author{Marcelo Guzm\'an, Denis Bartolo, David Carpentier}
\affiliation{Univ. Lyon, ENS de Lyon, Univ. Claude Bernard, CNRS, Laboratoire de Physique, F-69342, Lyon.}
\date{\today}

\begin{abstract}%
Systems as diverse as mechanical structures assembled from elastic components, and photonic metamaterials  enjoy a common geometrical feature: a sublattice symmetry. This property realizes a chiral symmetry first introduced to characterize
 a number of electronic insulators in the vicinity of their energy gaps.
{
In this article, we introduce a generic framework to elucidate and design  zero-energy topological boundary modes  in  all systems enjoying a chiral symmetry, whether crystalline or amorphous.
We first show how to distinguish chiral insulators  from one another by  a real-space  measure: their chiral polarization.
 In crystals,  we use  it to  redefine the very concept of  bulk-boundary correspondence, and  resolve long-standing ambiguities in its application to chiral insulators. 
  In amorphous metamaterials, we use it to lay out generic geometrical  rules  to locate topologically distinct phases, 
and explain how to engineer localized zero-mode wave guides even more robust than in periodic structures. }

\end{abstract}

\maketitle

{A century after the  foundations of band theory in  
solids by  F{\'e}lix Bloch~\cite{Bloch:1929}, physicists have discovered new 
states of electronic matter ranging from   insulators to  superconductors  by 
exploiting the  topological structure of Bloch theory ~\cite{Qi:2008,Hasan:2010,BernevigHughes,FranzMolenkampBook,AsbothOroszlanyaPalyi,Armitage:2018}.
\david{}{This topological revolution  has built on two cornerstones: 
 an
abstract classification 
based on symmetries}~\cite{schnyder2008classification,Kitaev2009,Ryu2010,Fidkowski:2011,Alexandradinata:2014,Taherinejad:2014,Alexandradinata:2016,Alexandradinata:2016b}, and the practical correspondence between bulk topology and the   boundary states measurable in experiments
~\cite{jackiw1976solitons,volkov1985two,fradkin1986physical,
Hatsugai:1993,Qi:2008,Hasan:2010,BernevigHughes,FranzMolenkampBook,AsbothOroszlanyaPalyi}.
During the past decade, these two generic principles spread 
frantically  across fields as diverse as photonics, acoustics, or mechanics, 
leading to design principles and practical realizations of  maximally robust 
waveguides~\cite{Ozawa2019,Mao2018}. }

Among the number of symmetries constraining wave topology, chiral symmetry has a special status.  Out of the three fundamental symmetries of the  overarching  ten-fold classification, it is the only one  naturally realized with all quantum and classical waves. 
It  generically takes the form of  a sub-lattice symmetry when  waves propagate in  frames composed of two connected lattices  $A$ and $B$, with 
 couplings only between, $A$ and $B$ sites, see e.g. Fig.~\ref{fig.2b}a.  In electronic systems, the archetypal example of a chiral insulator is provided by the polyacetylene molecule described by the Su-Schrieffer-Hegger (SSH) model~\cite{Su:1979}. In mechanics, the Hamiltonian description  of bead-and-spring networks is intrinsically chiral~\cite{Gurarie:2002,Gurarie:2003,Kane:2014,Huber2016}:  the $A$ sites correspond to the beads, and the $B$ sites to the springs.  In topological photonics and cold atoms chiral wave guides are among the simplest realizations of topological phases. 
\david{}{
Over the past decade,  the modern theory of electronic polarization based on Zak phases and non-Abelian Wilson loops~\cite{Zak:1989,Vanderbilt,Vanderbilt:1993,King-Smith:1993} 
 has illuminated the intimate relation between crystalline symmetries and the topology of band structures~\cite{Fidkowski:2011,Alexandradinata:2014,Taherinejad:2014,Alexandradinata:2016,Alexandradinata:2016b}. 
 By contrast, the role of chiral symmetry has been overlooked.
 
 In this article, by introducing the concept of chiral polarization we determine the zero-mode content of interfaces between topologically incompatible crystalline and amorphous chiral meta(materials) 
}
In the bulk, the chiral charge, which measures  the imbalance between the number of sites on the  
sub-frames
 $A$ and $B$, predicts the number of zero-energy modes of all Hamiltonians defined on a given chiral frame. 
 To characterize chiral insulators
we define their chiral  polarization $\bm \Pi$  
 as  the spatial imbalance of the bulk waves on the two sub-frames.
 This 
\david{}{
material property} does not rely on any crystalline 
 symmetry and can therefore be defined locally  
on disordered and  amorphous frames. In crystals, although akin to the time-reversal polarization of 
$\amsmathbb Z_2$ insulators~\cite{Fu:2006}, we show that $\bm \Pi$ is not merely set by the 
Bloch-Hamiltonian topology but also by the underlying frame geometry. At boundaries, we show how $\bm \Pi$ prescribes the surface chiral charge, and therefore the full 
zero-energy
edge content  of crystalline and amorphous chiral matter. 
    \david{}{Finally, we propose a series of practical protocols to  experimentally measure the chiral polarization of mechanical, and photonic chiral metamaterials.}

\section*{From chiral charge to chiral polarization  and  Zak phases}
\label{sec:ChargePolarization}
 Introducing the concepts of chiral charge and polarization, we demonstrate that bulk properties of chiral matter are determined by an intimate interplay between the frame topology, the frame geometry and the chiral Zak phases of Bloch Hamiltonians. \\
 
 \noindent
 {\bf{Chiral charge and chiral polarization.}}
 We consider the propagation of waves in  chiral material associated to  $d$-dimensional  frames including two 
sub-frames
 $A$ and $B$. The wave dynamics is defined by a  Hamiltonian $\mathcal H$. By definition, the chiral symmetry translates  in the anti-commutation of $\mathcal H$ with the chiral unitary operator 
$\amsmathbb{C}={\amsmathbb P}^A-{\amsmathbb P}^B$, where 
$\amsmathbb P^{A}$ and $\amsmathbb P^{B}$ are the two orthogonal projectors on the sub-frames $A$ and $B$. Simply put, in the chiral basis where $\amsmathbb C$ is diagonal, $\mathcal H$ is block off-diagonal. 
  
\begin{figure}[htb!]
\centerline{
	\includegraphics[width=\columnwidth]{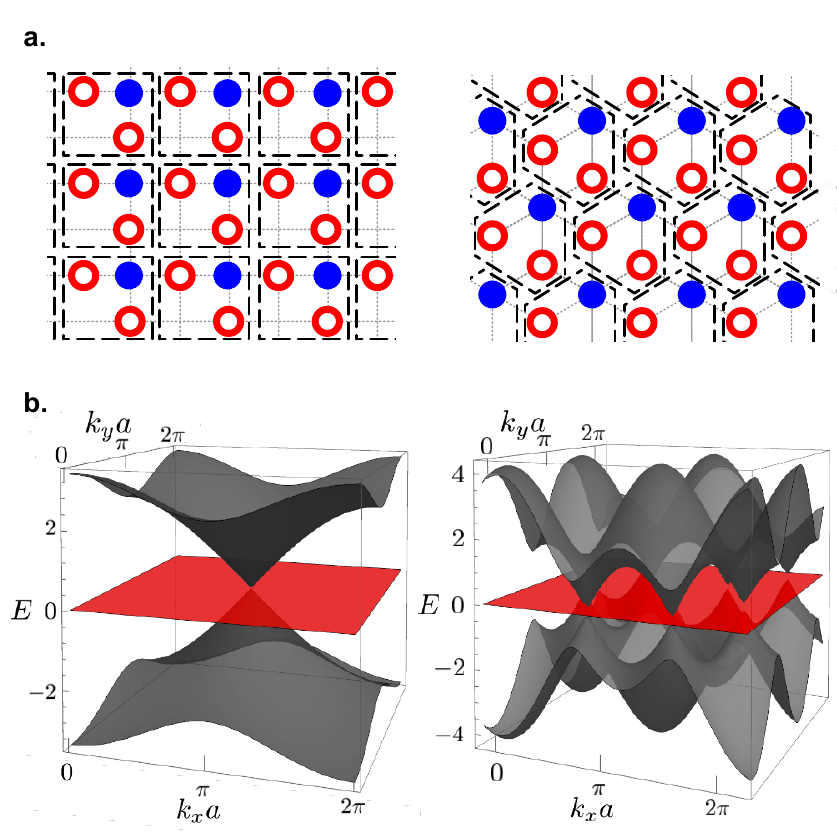}
	}
\caption{{\bf Lattices with a finite chiral charge}.
{\bf a.} The Lieb (left) and dice (right) frames are both characterized by an imbalance between the number $N^A$ and $N^B$ of sites. In both cases the chiral charge per unit cell equals 1. Any Hamiltonian defined on these frames possesses a flat energy band. {\bf b.} Illustration of two band spectra associated to chiral Hamiltonians defined on the  Lieb (left) and dice (right) frames. The two band spectra are computed for tight-binding Hamiltonians with nearest neighbour coupling and a hopping parameter set to 1, see e.g.~\cite{Louvet2015}.
}
\label{fig.2b}
\end{figure}
In order to  determine the relative weight of the wave functions of $\mathcal H$ on the two 
sub-frames, 
we introduce the chiral charge 
\begin{equation}
    {\mathcal M}=  \langle \amsmathbb C\rangle,
    \label{eq:defM}
\end{equation}
where the average is taken over the complete Hilbert space. Using the basis of fully localized states, we readily find  that $\mathcal M$ is fully prescribed 
by the frame topology: the chiral charge counts  the imbalance  between the number of $A$ and $B$ sites:   $ {\mathcal M}=N^A-N^B$ . We can however also 
evaluate Eq.~\eqref{eq:defM} in the eigenbasis of $\mathcal H$. Indexing by $n$ the 
eigenenergies
 of $\mathcal H$, the eigenstates of the chiral Hamiltonian 
come by pairs of opposite energies related by $\ket{-n}=\amsmathbb C\ket{n}$. Chirality therefore implies that  the chiral charge is solely determined by the 
zero modes of $\mathcal H$ as $\mathcal M=\sum_n\bra{n}\amsmathbb C\ket{n}=\bra{0}\amsmathbb C\ket{0}$. Noting that the $\ket{0}$ states are  eigenstates of 
the chiral operator with eigenvalue $+1$ when localized on the $A$ sites and $-1$ when localized on the B sites, it follows that $\mathcal M$  also is an 
algebraic count the zero modes of $\mathcal H$:
\begin{equation}
{\mathcal M}=N^A-N^B=\nu^A-\nu^B.
\label{eq:MaxwellCalladine}
\end{equation}  
This equality is the classical result established by Maxwell and Calladine in the context of structural mechanics~\cite{Maxwell:1864,Calladine:1978} and independently discussed by Sutherland in the context of electron localization \cite{Sutherland1986}. 
Eq.~\eqref{eq:MaxwellCalladine} implies that the spectral properties of $\mathcal H$ are constrained by the frame topology. 
In particular, frames with a non-vanishing chiral charge impose {\em all} chiral Hamiltonian to possess flat bands.
This simple prediction is illustrated in Fig.~\ref{fig.2b} where we show the Lieb and the dice lattices, which are both characterized by a unit chiral charge per unit cell. 
All Hamiltonians defined on these  lattices are therefore bound to support at least one flat band, Fig.~\ref{fig.2b}b. 
No chiral insulators exist on the Lieb and dice lattices.

By contrast, in chiral insulators, no zero-energy bulk modes exist and   $\mathcal M$  must vanish. To probe the relative weight of the wave functions on the 
two 
sub-frames, 
we therefore introduce the chiral polarization vector 
    {${\Pi}_j=   \langle \amsmathbb C x_j\rangle_{E\neq 0}$.}
 As the $\ket{\pm n}$ states  contribute equally to $\Pi$  in chiral systems,  we henceforth use the definition
  \begin{equation}
 {\Pi}_j= 2 \langle \amsmathbb C x_j\rangle_{E<0},
 \label{eq:defPolarization}
 \end{equation}
with $j=1,\ldots,d$ are the indices of the $d$ crystallographic directions and  where $E<0$ indicates that the  average is taken over the occupied states.
\david{}{Although  seemingly identical to the the skew polarization 
introduced in ~\cite{MondragonShem:2014,Rakovszky:2017} for topological insulators, and the mean chiral displacement of quantum walks~\cite{Cardano:2017}, we emphasize that ${\Pi}_j$  does not rely on any Bloch representation and is therefore defined also in amorphous phases. 
We stress that, even in the crystalline  case,  ${\Pi}_j$ includes  content 
beyond the skew polarization,
as it resolves the weighted positions with a sub-unit-cell resolution.
These  differences are not mere technicalities, and will prove crucial in the next sections.} 

\begin{figure}
\begin{center}
	\includegraphics[width=8.5cm]{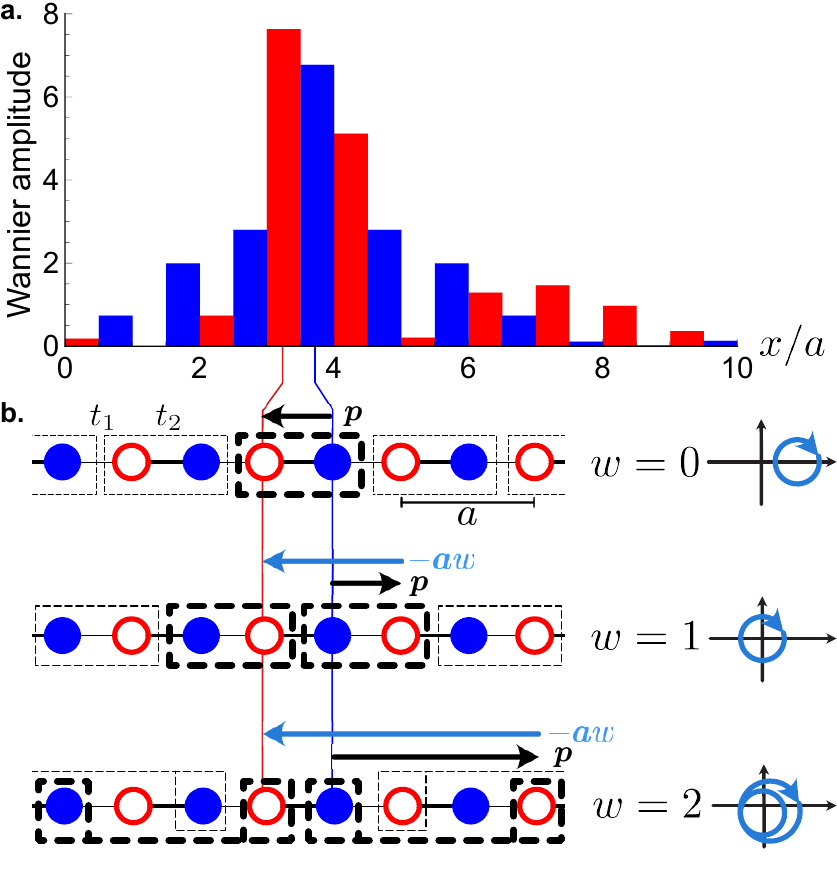}
\end{center}
\caption{{\bf Chiral polarization and Wannier functions.}
{\bf a.} Square of the Wannier amplitude projected into the $A$ (red) and $B$ (blue) sublattices for the ground state configuration of the two-band SSH model as defined in~\cite{SSHWANNIER}, with hopping ratio $t_1/t_2=0.79$. $a$ denotes the period of the 1D frame. {The chiral polarization $\Pi=\expval{x^A}-\expval{x^B}$ is negative: the chain is left polarized regardless of the choice of unit cell.} {\bf b.} The  winding number of the Bloch Hamiltonian encodes the chiral polarization \emph{relative} to a given unit cell. The chiral polarization being a material property, the winding number $w$ can therefore take any integer value when redefining the geometry of the unit cell as illustrated in the last column. Whatever the choice of the unit cell, the difference between the geometrical polarization  and $aw$ has a constant value given by the chiral polarization $\Pi$.}
\label{fig.2}
\end{figure}

To gain more physical insight, it may be worth noting that in 
 electronic systems, $\Pi_j$ corresponds to the algebraic distance between the charge centers associated to the $A$ and $B$ atoms. While in mechanical networks,  $\Pi_j$ is the vector connecting the stress-weighted and displacement-weighted positions.
A vanishing polarization indicates that the average locations of the stress and displacement coincide. Conversely, a finite chiral polarization  reveals an asymmetric mechanical response 
discussed in~\cite{Rocklin2017,Bilal2017}.
For the sake of clarity, before  revealing  topologically protected zero modes in amorphous phases,  we first explore the consequences of a 
finite chiral polarization in periodic systems such as in the paradigmatic example of the SSH model illustrated Fig.~\ref{fig.2}. 
\\

\noindent
{\bf{Chiral polarization: an interplay between  Zak phases and frame geometry.}}

We begin with a thorough discussion of crystalline materials, defined by periodic frames and  Bloch Hamiltonians.
\david{}{Building on previous works on the electronic polarization~
\cite{Zak:1989,Vanderbilt,Vanderbilt:1993,King-Smith:1993}, 
}
we relate the chiral polarization of a crystalline material 
to the two Zak phases of waves projected on sub-lattices $A$ and $B$ {when  transported across the Brillouin zone}.  
To do so, we  first  choose a unit cell and consider the basis of Bloch states  
$\ket{{\bm k},\alpha}=\sum_{\bm R} e^{i{\bm k}\cdot \bm R}\ket{\bm R+\bm r_{\alpha}}$, where 
$\bm R$ is a Bravais lattice vector, $\alpha$ labels the  atoms in the unit cell and ${\bm k}$ is the momentum in 
the Brillouin Zone (BZ). We henceforth use a convention where  the Bloch Hamiltonian $H({{\bm k}})$ is periodic in the BZ, see~\cite{Blount:1962,Vanderbilt} and Methods. %
 More quantitatively, considering first Hamiltonians with no band crossing~\footnote{%
In the situation where bands cross, our results should be generalized resorting to the Wilson loops of the non-commutative Berry connexion instead of the abelian Zak phase connection~\cite{Neupert:2018}.}, we define the $A$ sub-lattice  Zak phase
of the $n^{\rm th}$ energy band along the crystallographic direction $j$ as 
\begin{equation}
\gamma^{A}_{j}(n)=i\int_{\mathcal C_j} d{\bm k} 
\expval{ {\amsmathbb P}^{A}\partial_{\bm k}{\amsmathbb P}^{A} }{u_n},  
\end{equation} 
where the $\ket{u_n({\bf k})}$  are the eigenstates of $H({\bm k})$,  and  $\mathcal{C}_j$ the  non-contractible loops  over the Brillouin zone defined along  the $d$ 
crystallographic axes. $\gamma^{B}_{j}(n)$ is defined analogously on the $B$ sublattice. %
The (intercellular) Zak phase is given by the sum of $\gamma^{A}_j(n)$ and $\gamma^{B}_j(n)$~\cite{Rhim:2017}.  In Methods, we show how to decompose the chiral polarization into a spectral and a frame contribution:
\begin{align}
	\Pi_j =  \frac{a}{\pi}(\gamma^{A}_j-\gamma^{B}_j)+  p_j,
\label{eq:defPi}
\end{align}
where $a$ is the lattice spacing (assumed identical in all directions), $\gamma^{A}_{j}$ and $\gamma^{B}_{j}$ are the  sublattice Zak phases defined by
\begin{equation}
\gamma^{A}_{j}=\sum_{n<0}\gamma^{A}_{j}(n).
\end{equation}
In Eq.~\eqref{eq:defPi} the $p_j$ are the components of the geometrical-polarization vector  connecting the centers of mass of the $A$ and $B$ sites in the unit-cell:
\begin{align}
\bm p= \sum_{\alpha\in A}\bm r_{\alpha}-\sum_{\alpha\in B}\bm r
_{\alpha}.
\label{eq:defpi}
\end{align}
In  crystals, Eqs.~\ref{eq:defPi} quantifies the difference between the polarity of the ground-state wave function $\bm \Pi$ and the geometric polarization of the frame $\bm p$. This difference is finite only when the two sublattice Zak phases differ.
\section*{Topology of chiral insulators}
\label{sec:Topology}
\begin{figure*}
\centerline{	\includegraphics[width=\textwidth]{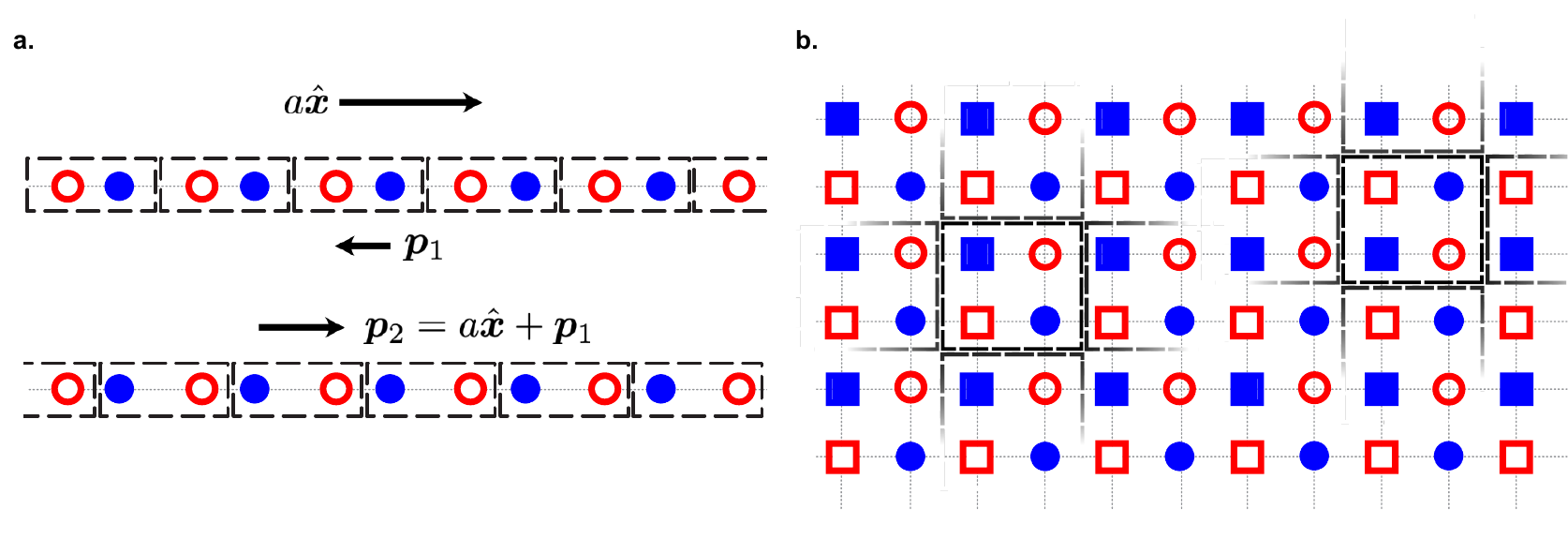}
}
	\caption{{\bf Inferring the band topology from frame geometry}.
	{\bf a.} The two-sites Wigner-Seitz cell on a 1D chiral frame have different geometrical polarizations; their difference is given by one Bravais vector. Consequently, we can always define the unit cell so that the Bloch Hamiltonian has a finite winding.
	{\bf b.} All the Wigner-Seitz unit cells on the checkerboard lattice  share the same (vanishing) chiral polarization. Therefore a single winding number $w$ characterizes the Hamiltonians on this frame in virtue of Eq.~\eqref{eq:indexP}. Evaluating the winding using the Wigner-Seitz cell compatible with the atomic limit of $\mathcal H$ yields $w=0$, by definition.
}
\label{fig.4}
\end{figure*}
We now elucidate the intimate relation between  the chiral polarization and the band topology  of chiral gapped phases 
{defined on periodic lattices}. We outline the demonstrations of our central results below and detail them 
in Methods.\\

\noindent
{\bf{Sublattice Zak phases and winding numbers.}}
Computing the Wilson loop of the non-Abelian connection  $\mathbf{A}_{n,m}({\bm k}) = 
    \bra{u_n ({\bm k}) }\partial_{\bm k} \ket{u_m({\bm k})}$ along $\mathcal C_j$, we  show that chirality relates the $d$ Zak phases $\gamma_j^{A}+\gamma_j^{B}$ to the windings of the Bloch Hamiltonian as 
\begin{equation}
\gamma_j^A+\gamma_j^B=\pi w_j\,+2\pi\amsmathbb Z,
\label{eq:gammaApgammaB}
\end{equation}
where $w_j=i/(4  \pi)\int_{\mathcal C_j} d{\bm k}\cdot \Tr [\partial_{\bm k} H\amsmathbb{C}H^{-1} ]\in{\amsmathbb Z}$. The total Zak phase is quantized but the arbitrary choice of the origin of space implies that both $\gamma^{A}$ and $\gamma^{B}$ are only defined up to an integer. As a matter of fact, a mere $U(1)$ gauge transformation $\ket{u_n}\rightarrow e^{i\alpha_n (k)}\ket{u_n}$  arbitrarily modifies  $\gamma_j^{A}(n)$ and $\gamma_j^{B}(n)$ by the same quantized value:  $\gamma_j^{A}(n)\to \gamma_j^{A}(n)+\pi m$, $\gamma_j^{B}(n)\to \gamma_j^{B}(n)+\pi m$, with $m\in \amsmathbb Z$. 
By contrast, the difference between the two sublattice Zak phases  is  left unchanged by the same gauge transformation  which  echoes its independence from the space origin. Evaluating the winding of $H(\bm k)$ using the Bloch eigenstates (see Methods), we readily establish the essential  relation\footnote{Note that this difference of Zak phases was recently denoted as a chiral phase index in \cite{jiang2020topology}.}
\beq
\gamma^B_j - \gamma^A_j=\pi w_j\;\;\;\in  \pi\amsmathbb{Z}.
\label{eq:defw}
\eeq
Chirality quantizes the  sublattice Zak phases of chiral insulators, even in the absence of  inversion or any other specific crystal  symmetry. $\gamma_j^A$ and $\gamma_j^B$ are however not independent. Combining  Eqs.~\eqref{eq:gammaApgammaB} and ~\eqref{eq:defw}  we can always define the origin of space so that 
$\gamma^A_j=0$ and $\gamma_j^B=\pi w_j$. 

The $d$  winding numbers of Eq.~\eqref{eq:defw} characterize the topology of $H({{\bm k}})$. In particular,  if  for a given Wigner-Seitz  cell the corresponding $H(\bm k)$ is associated to a finite winding  ($w_j\neq0$), then it cannot be smoothly deformed into the atomic limit defined over the same unit cell. 
The set of winding numbers  is however  poorly informative about the spatial distribution of the charges in electronic systems, or about the stress and displacement distributions in mechanical structures. The values of $w_j$ are  defined  only up to the arbitrary choice of unit cell required to construct the Bloch theory. A well known example of this limitation is given by the SSH model, where the winding of $H_{{\bm k}}$ can either take the values 
0 or $\pm1$ depending on whether the unit cell's leftmost site belongs to the $A$ or $B$ sublattice, 
see Fig.~\ref{fig.2}a and Methods. 
We show in the next section, how the chiral polarization alleviates this limitation.\\

\noindent
{\bf{Disentangling   Hamiltonian topology from frame geometry.}} Equations \eqref{eq:defPi} and \eqref{eq:defw}   provide a clear geometrical interpretation of the winding number $w_j$ as 
 the quantized difference between the geometrical and the chiral polarization:
\begin{align}
\Pi_j = \left(p_j  - {a_j} w_j\right ).
\label{eq:polarizationDecomp}
\end{align}
We can now use this relation to illuminate the very definition of a chiral topological insulator. The chiral polarization 
$\Pi_j=2 \langle\amsmathbb C x_j\rangle_{E<0}$ is a physical quantity that does not depend on the specifics of the Bloch representation. Therefore computing $\Pi_j$ for two unit cells $(1)$ and $(2)$, we  find that the windings of the two corresponding Bloch Hamiltonians $ H^{(1)}({\bm k})$ and $ H^{(2)}({\bm k})$ are related via Eq.~\eqref{eq:polarizationDecomp} as
 \begin{equation}
w^{(2)}_{j} - w^{(1)}_{j}   =\frac{1}{a_j} \left( p^{(2)}_{j} - p^{(1)}_{j}\right ).%
\label{eq:indexP}
\end{equation}  
This essential relation implies that one can always construct a Bloch representation of $\mathcal H$ where $H(\bm k)$ is topologically trivial, at the expense of a suitable choice of a unit cell. As a matter of fact, a redefinition of the unit cell can increase, or reduce the geometrical polarization, and therefore the winding numbers, by an arbitrary large multiple of $a_j$ as illustrated in Fig.~\ref{fig.2}b.

For instance in the case of Hamiltonians with nearest neighbor couplings, applying Eq.~\eqref{eq:indexP} to Wigner Seitz unit cells ($|w_j|\leq1$), we find that there exist as many topological classes of
$\mathcal H$, as different geometrical polarizations in the Wigner-Seitz cells. This number provides a
direct count of the chiral 'atomic limits' of $\mathcal H$.

 Defining the  topology of a chiral material  therefore requires characterizing both the winding of its
 Bloch Hamiltonian, and the frame geometry. Remarkably, this interplay provides an insight on topological
 band properties from the sole inspection of the frame structure.\\
 
\begin{figure*}
\begin{center}
	\includegraphics[width=\textwidth]{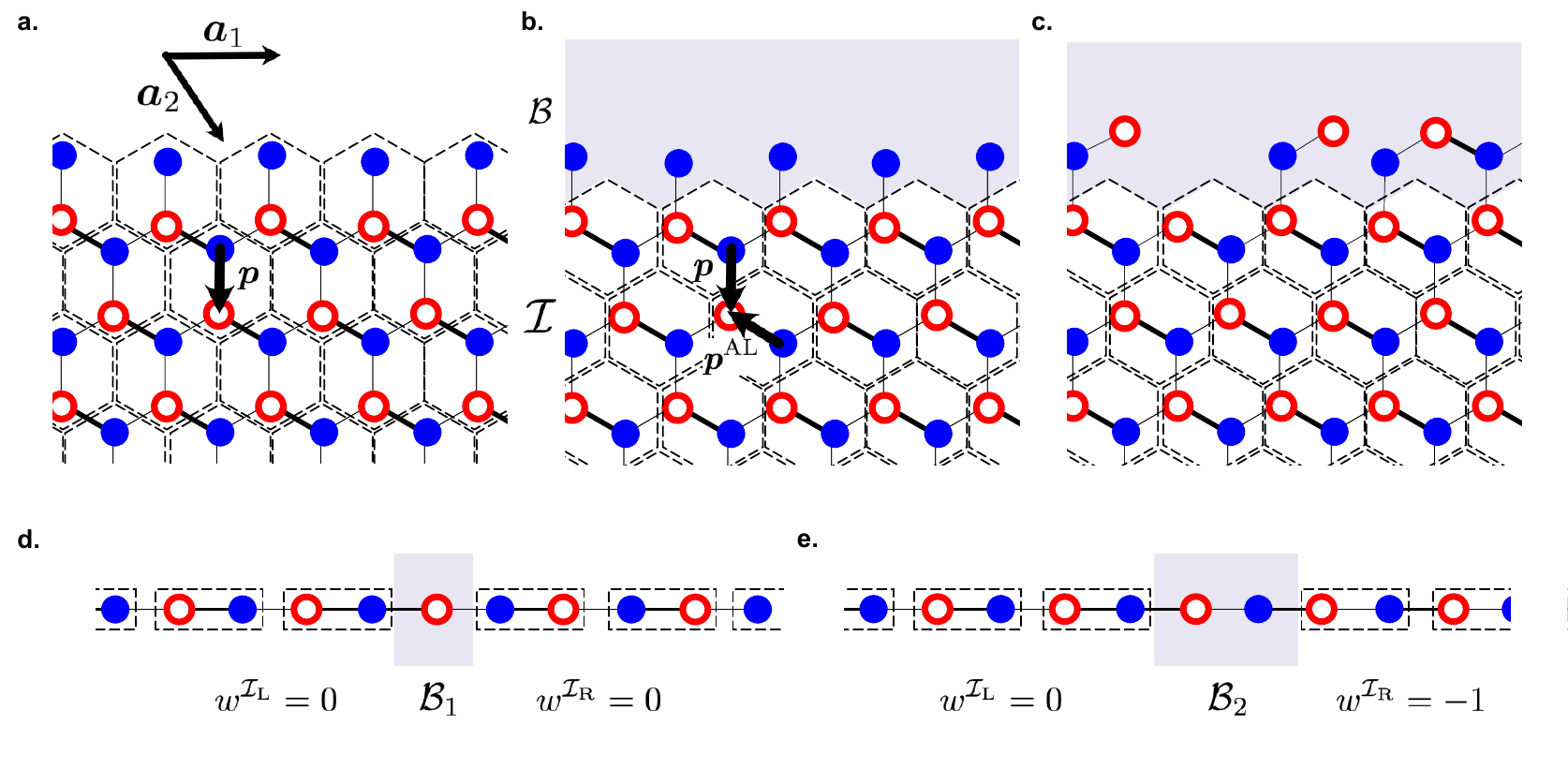}
\end{center}
\caption{
{\bf Bulk-boundary correspondance.}
\david{}{ 
{\textbf a.} A chiral crystal defined on a honeycomb frame is terminated by a clean zigzag edge {incompatible} with the atomic-limit Hamiltonian defined by keeping only the dominant couplings represented by thick solid lines. 
The dashed rectangles indicate the Wigner-Seitz cells allowing a tessellation compatible with {the edge geometry}. The arrow indicates the geometrical polarization {$\bm p$}. 
{\textbf b.} Same physical system. The crystalline bulk is now tiled using {the unit cell compatible with the atomic limit}. This requires a redefinition of the crystal boundary {$\mathcal B$} (shaded region). The arrows indicate the geometrical polarization of the new unit cell {($\bm p^{\rm AL}$)}. The difference $\bm p-\bm p^{\rm AL}$ is a Bravais lattice vector ($\bm a_2$).
{\textbf c.} Same material as in (\textbf a.) and (\textbf b.)  including a disordered interface $\mathcal B$ bearing a non-zero chiral charge $\mathcal M^{\mathcal B}$.
{\textbf d.} Two connected SSH chains. The Wigner-Seitz cell in the two materials are compatible with their atomic limits.  The interface $\mathcal B_1$ separating the two materials is one-site wide.
{\textbf e.} Redefining the Wigner-Seitz cell on the right hand side of the interface requires widening the boundary region. This redefinition makes the unit cell incompatible with the atomic limit. The winding of the Bloch Hamiltonian in $\mathcal I_{\rm R}$ takes a finite value and consequently modifies the zero-mode content of the boundary region.
}
}
\label{fig.3}
\end{figure*}
\noindent
{\bf{Inferring band topology from  frame geometry.}} There exists no trivial chiral phase  in one dimension: one can always choose a Wigner-Seitz cell such that the 
Bloch representation of $\mathcal H$  has a non-vanishing winding. As a matter of fact, the geometrical polarization of the Wigner-Seitz cells  can only take two finite values of opposite sign depending on whether the leftmost site in a unit cell is of the $A$ or $B$ type, see Fig.~\ref{fig.4}a. %
Equation~\eqref{eq:indexP} therefore implies that, in $1D$, there always exists, at least, two topologically distinct gapped phases smoothly connected to two atomic limits. The two gapped phases are  characterized by  two distinct pairs of winding numbers defined by two inequivalent choices of  unit cells. In other words {\rm all} SSH Hamiltonians are topological.

 Similarly, in $d>1$  only  frames having a geometrical polarization invariant upon redefinition of the Wigner-Seitz cell can support  topologically trivial Hamiltonians.
Equation~\eqref{eq:indexP} indeed implies that  a topologically trivial Hamiltonian $\mathcal H$ 
constrains the frame geometry to obey  $ p^{(1)}_{j} = p^{(2)}_{j}$ for all pairs of unit cells and in all directions $j$. 
We show a concrete  example of such a frame in  Fig.~\ref{fig.4}b. 
 
Before discussing the crucial role of the frame topology and  geometry on the bulk-boundary correspondence of chiral phases, we extend these two notions to chiral insulators with a flat band. \\

\noindent \david{}{
{\bf Chiral polarization in the presence of a net chiral charge. }
It is worth noting that the chiral polarization can also be defined and computed in the presence of an additional zero-energy flat band in the gap. As detailed in the Methods section, it then takes the form  
\beq
{\Pi}_j= ( p_j -p^{\text{ZM}}_j)+
  a \left(\bm{\gamma}_j^A-\bm{\gamma}_j^B\right)/\pi.
  \label{eq:PiZM}
\eeq
In this case, we loose the clear decomposition $\Pi$ into geometrical and topological contributions.
The geometrical polarization is corrected by 
$\bm p^{\text{ZM}}$  which originates from a spectral contribution associated to the zero-energy band. Furthermore the second term on the r.h.s., the difference between two
geometrical Zak phases, is not a topological winding number anymore. Despite the seemingly  complex form of Eq.~\eqref{eq:PiZM}, we  show in the next section that the chiral polarization remains an effective tool to relate
spectral bulk properties to the number of zero-energy states localized at boundaries.}

\section*{Bulk-boundary correspondence}
\label{sec:BulkBoundary}

\noindent
{\bf{Topological chiral charge of surfaces and interfaces.}
\label{Sec:BB}}
We now establish a bulk-boundary correspondence relating the chiral polarization  to the number of zero modes supported by the free surface of a chiral insulator. For the sake of clarity, we discuss the two-dimensional case without loss of generality.  We  consider first a  crystalline insulator 
$\mathcal I$  terminated by a clean edge $\partial \mathcal I$ oriented along a Bravais vector, say $\bm a_1$  as illustrated in~Fig.~\ref{fig.3}a. 

{
\david{}{
The bulk of the insulator can be described by different types of unit cells. As illustrated in Fig.~\ref{fig.3}a, in the presence of a clean edge, it is natural to choose a unit cell which allows a  tessellation of the whole system. 
However, this unit cell is generically incompatible with the atomic limit of the  Hamiltonian, and therefore does not allow a direct count of the zero energy boundary states using the simple Maxwell-Calladine count. An obvious strategy hence consist in redefining the unit cell, as in Fig.~\ref{fig.3}b to match the constraints of the atomic limit. This redefinition comes at the expense of leaving sites outside of the bulk tessellation. We define this ensemble of sites as the boundary region $\mathcal B$. Keeping in mind that we can smoothly deform the Hamiltonian into its atomic limit without closing the gap, we use Eq.~\eqref{eq:MaxwellCalladine} to count the number of zero energy states hosted by $\mathcal B$. It is given by 
${\mathcal V}={\mathcal M}^{\mathcal B}$. An essential geometrical observation is that the net chiral charge in $\mathcal B$ can be expressed as    
${\mathcal N^{\partial\mathcal I}}( p_2^{\rm AL}- p_2)$, where ${\mathcal N^{\partial\mathcal I}}$ is the edge length expressed in number of unit cells and $p_2$ is the geometrical polarization of the initial unit cell. We can now make use of the invariance of the chiral polarization formalized by Eq.~\eqref{eq:indexP} to relate the geometrical count of zeromodes to the winding of the Bloch Hamiltonian:
$\mathcal V={\mathcal N^{\partial\mathcal I}}(p_2^{\rm AL}-p_2)={\mathcal N^{\partial\mathcal I}}w^{\mathcal I}_2$. To arrive at a bulk boundary correspondence generic to all chiral insulators, we include the possibility of dealing with irregular interfaces featuring a net chiral charge $\mathcal M^{\mathcal B}$ as sketched in Fig.~\ref{fig.3}c. We then find 
\begin{equation}
\mathcal V
= \mathcal{M}^{\mathcal B}+{\mathcal N^{\partial\mathcal I}}~ {w}^{\mathcal I}_2.
\label{eq:KaneLubensky}
\end{equation}
Three  comments are in order. }
Firstly, the bulk boundary correspondence  defined by Eq.~\eqref{eq:KaneLubensky}
illuminates the geometrical implication of a  nonzero winding: a finite $w_j^{\mathcal I}$  echoes the impossibility to tile a periodic frame with unit cells compatible with the Hamiltonian's atomic limit.
Secondly, 
Eq.~\eqref{eq:KaneLubensky} is readily generalized to  interfaces separating two chiral insulators $\mathcal I_{\rm L}$ and $\mathcal I_{\rm R}$, where we simply have to apply the same reasoning on each side of the interface: $\mathcal V
= \mathcal{M^B}+{\mathcal N}^{\partial{ \mathcal I}}({w}^{\mathcal I_{\rm L}}+{ w}^{\mathcal I_{\rm R}})$, see e.g. Figs.~\ref{fig.3}d and~\ref{fig.3}e. 
Thirdly, the formula given by Eq.~\eqref{eq:KaneLubensky} generalizes the Kane-Lubensky index introduced in their seminal work to count the zero-energy modes localized within isostatic mechanical networks~\cite{Kane:2014}. %
We  show  that this index defines a bulk-boundary correspondence generic to all chiral insulators and even to flat band insulators such as hyperstatic lattices as further discussed in the Methods section.

\section*{Amorphous Chiral Insulators}
\label{Amoprhe}
\begin{figure*}[ht]
\begin{center}
	\includegraphics[width=\textwidth]{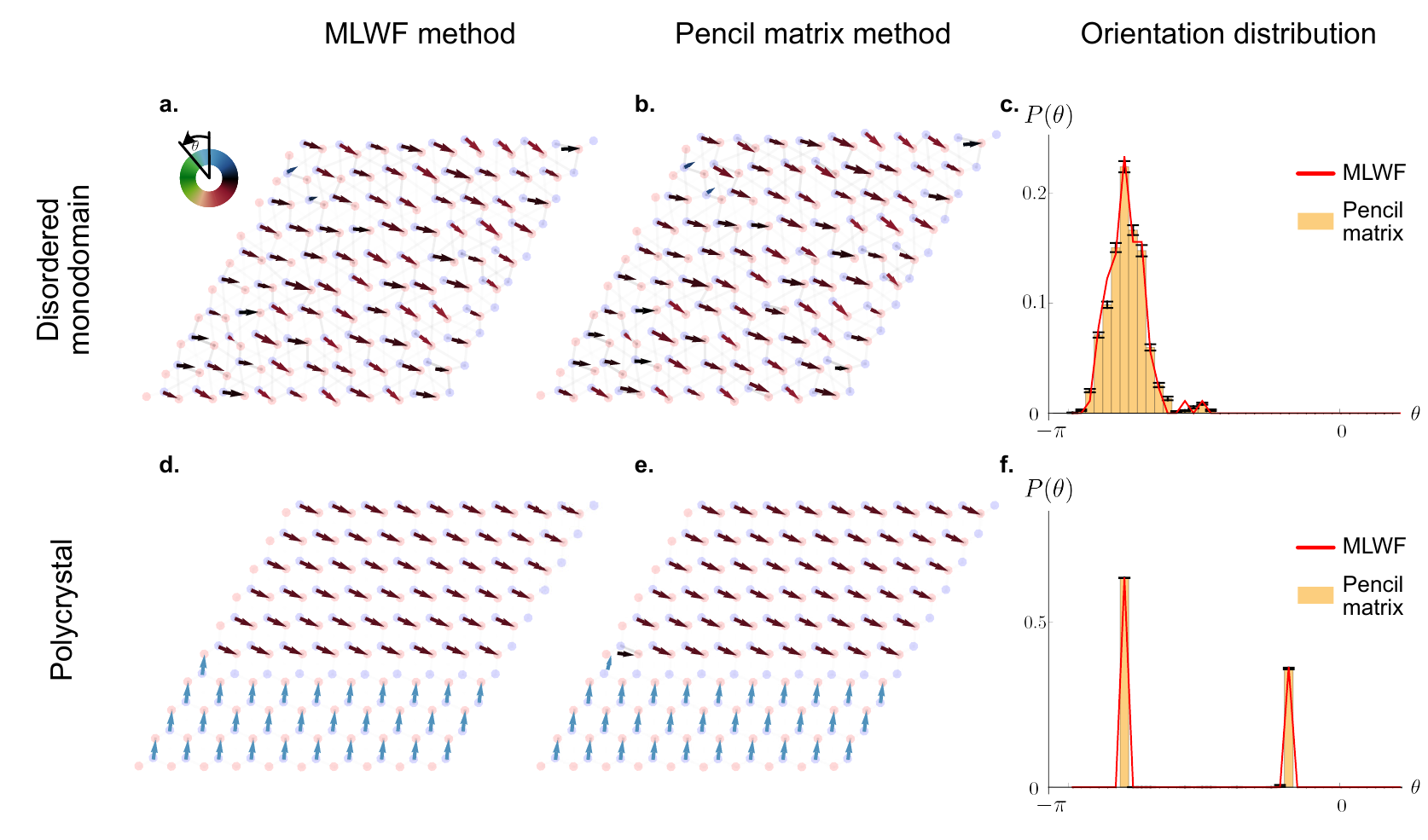}
\end{center}
\caption{
\david{}{
{\bf Pencil matrix versus maximally localized Wannier functions}
{\bf a.} Single domain configuration with geometrical and spectral disorder. The chiral polarization field obtained from the maximally localized wannier functions is superposed. {\bf b.} Chiral polarization field obtained from one realization of the pencil matrix procedure. {\bf c.} Orientation distribution obtained from 50 values of $\alpha$ (bar chart), and from the maximally localized wannier functions (red solid line). {\bf d.}, {\bf e.}, {\bf f.} correspond to the same information, this time for two crystalline domains.  
}
}
\label{fig.6.5}
\end{figure*}

{In  condensed matter, chiral symmetry is a low energy feature of electronic Hamiltonians, which is unlikely to survive to strong structural disorder. Conversely, in photonic, accoustic or mechanical metamaterials chirality is built in by design and can therefore be present both in ordered or amorphous structures~\cite{Ozawa2019,Zhang201}. In mechanical metamaterials chirality is even more robust as it is inherent to any system assembled from elastically coupled degrees of freedoms~\cite{Mao2018}. In this section, we show how to generalize our physical characterization of zero energy modes to disordered  chiral  metamaterials.

 Over the past two years a number of experimental, numerical and theoretical works showed that crystalline symmetries are not required to define topological insulators, 
 see e.g. ~\cite{Mitchell2018,Xiao2017,Agarwala2017,Marsal2020}. Unlike these pionneering studies where 
 topologically inequivalent disordered
 phases 
 are distinguished by
 abstract indices defined in real space 
 and 
 related to the quantification of edge currents, our framework solely based on the chiral polarization applies to chiral systems regardless of the presence or not of  time reversal 
 symmetry. }

Our strategy  follows from the fundamental relation: $  \Pi_j=p_j-a_jw_j$ of Eq.~\eqref{eq:polarizationDecomp}. This relation  implies a one-to-one correspondence between  the chiral polarization and 
a topological spectral property quantized by
 the winding vector.  
The basic idea hence consists in probing the existence of topologically protected zero modes by {\em local} discontinuities in the chiral polarization field, even when no winding number or Zak phase can be defined. Relating topologically protected excitations to real-space singularities requires defining a local chiral polarization field $\bm \Pi(\bm x)$. 
By definition, 
$\bm \Pi(\bm x)$  measures  the local imbalance of 
the wave function 
carried by the $A$ and $B$ 
sites. 
\david{}{
To express $\bm \Pi(\bm x)$,  it would be   natural to consider
eigenstates  of the position operator $P\bm{x}P$ 
projected onto the occupied states of $\mathcal H$. 
However, in dimension $d>1$, the different components of the projected position operator do not commute $[Px_jP,Px_kP]\neq 0$ for $j\neq k$, and do not possess common eigenstates. Instead, we express the 
polarisation in terms of the 
maximally localized states ${\widetilde W_m}$ \cite{Vanderbilt}, 
which are 
centered  on the position 
 $\bm x_m \equiv \expval{\widehat X}{\widetilde W_m}$. These states  
generalize the Wannier functions in the absence of translational symmetry, see Methods %
for more details. 
We can then define the {\em local} chiral polarization  as the weighted chiral position evaluated over $\widetilde W_m$:
\beq
\bm{\Pi}(\bm x_m) =2\expval{\amsmathbb C \widehat X}{\widetilde W_m}.
\label{eq.chiralPolGeneralText}
\eeq

In practice, we can bypass the time consuming numerical determination of the ${\widetilde W_m}$ by taking advantge of the so-called pencil-matrix 
method~\cite{golub2013matrix}. In short, the method consists  in replacing in 
\eqref{eq.chiralPolGeneralText} 
the ${\widetilde W_m}$ by eigenstates of a linear 
combination of the projected position components 
$ R = \sum_j \alpha_j P x_j P\ ;\  \sum \alpha_j = 1 $. 
The  dependence on $\alpha_i$ of the resulting chiral polarization 
is a  measure of the non-commutativity of the $Px_j$ typically associated to a nonvanishing Berry curvature. 
In practice, as illustrated in Fig.~\ref{fig.6.5}, the difference between the actual polarization, computed from the ${\widetilde W_m}$, and its approximation based on the $R$-matrix eigenstates is smaller than the distance between neighboring sites. 
Given the excellent agreement found both in mono and polycrystals, we henceforth use the pencil matrix method to locally measure the chiral polarization fields in disordered and amorphous structures out of reach of conventional chiral displacement characterizations
\cite{meier2018observation}. 
}

To make the discussion as clear as possible we  
 consider separately the 
two possible sources of randomness in a disordered chiral insulator:
(i) geometrical disorder, which affects  the frame geometry leaving the interaction between the $A$ and $B$ sites unchanged and  (ii) 
Spectral 
disorder, which alters the interactions while living the frame geometry unchanged.\\

\begin{figure*}[ht]
\begin{center}
	\includegraphics[width=\textwidth]{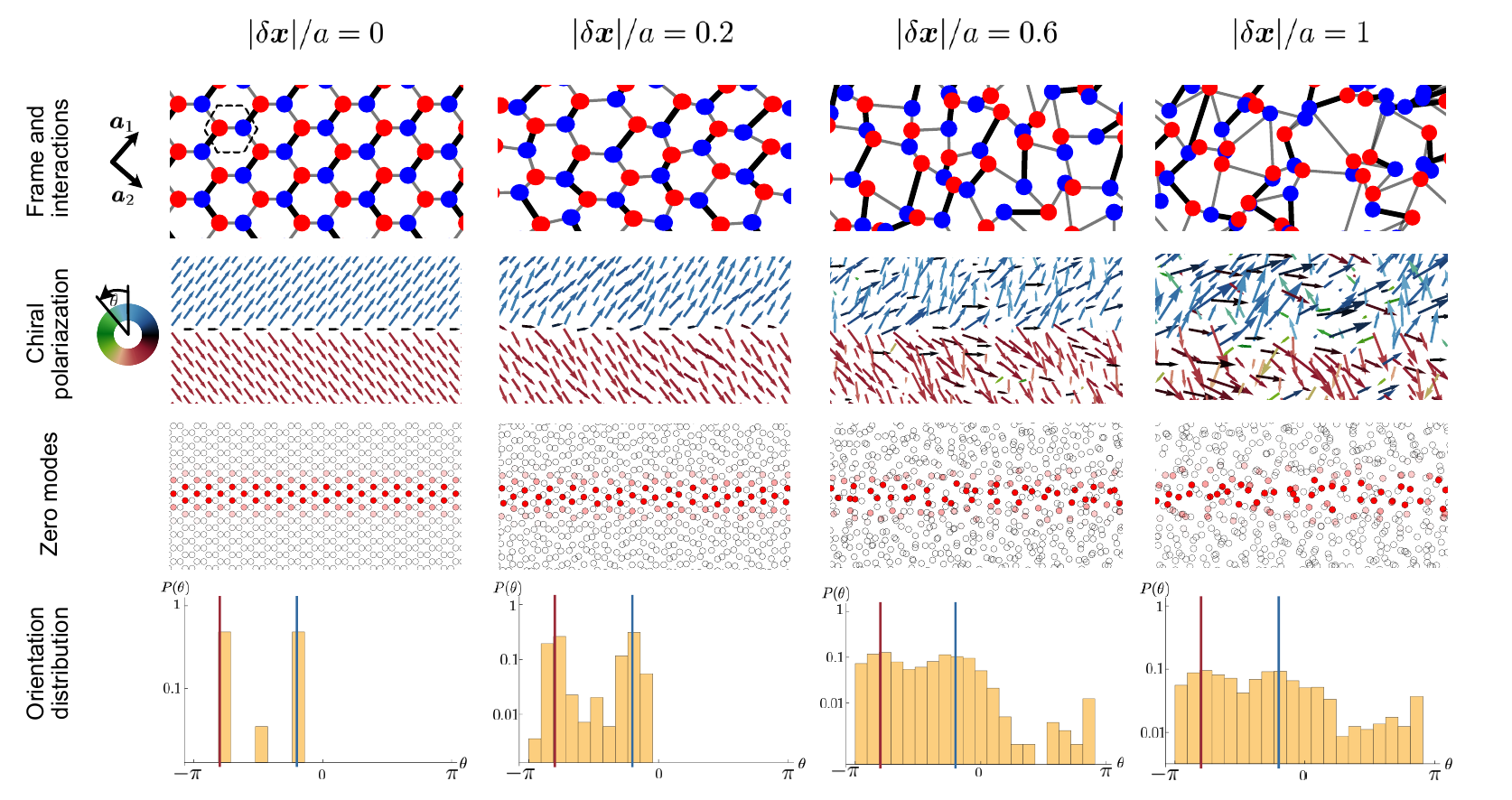}
\end{center}
\caption{
{\bf Topological zero energy states  on amorphous frames}
First row: Sketch of the frame geometry for increasing positional disorder quantified by the maximal amplitude of the random displacements $|\delta \bm x|/a$.  All panels show the vicinity of a boundary between two different insulators defined on the same frame but with different positions of the stronger couplings. 
The lines' width indicates the magnitude of the coupling strength. In all panels $t'/t=20$. In the lefmost panel, we indicate the choice of the unit cell and of the crystallographic axes.
Second row:  Corresponding chiral polarization fields. The color indicates the orientation of $\bm \Pi(\bm x)$ 
Third row:  Magnitude of the zero-mode wave function. The zero mode is located at the boundary between topologically inequivalent states even on amorphous frames.
Fourth row: Probability density function of the $\theta$, the local orientation of the chiral polarization field. The distributions are peaked on the same two directions (vertical lines) regardless of the magnitude of disorder. This reveals the coexistence of two distinct topological phases robust to positional disorder. 
}
\label{fig.5}
\end{figure*}

\noindent
{\bf{Topological zero modes on amorphous chiral frames.}}
The reasoning is easily explained starting from a concrete example.
Fig.~\ref{fig.5} shows the interface between two topologically distinct insulators,
$\mathcal I_{\rm T}$ and $\mathcal I_{\rm B}$, living on a honeycomb frame.
They correspond to  distinct atomic limits of a nearest-neighbor tight binding Hamiltonians including two different hopping coefficients, see e.g.~\cite{Bellec:2013}. For the choice of unit cell sketched in Fig.~\ref{fig.5}, the winding vectors are 
$\bm w^{{\mathcal I}_{\rm T}}=(0,1)$ and $\bm w^{{\mathcal I}_{\rm B}}=(1,0)$.
As a result the boundary region $\mathcal B$ hosts one zero mode per unit cell located on the $A$ sites. As expected from Eq.~\eqref{eq:polarizationDecomp}, on a homogeneous periodic frame,  $\bm \Pi(\bm x)$ 
takes two distinct values in the two regions, and 
is discontinuous across $\mathcal B$.
Correspondingly, 
the distribution of the chiral polarization 
in the sample consists of two peaks centered on the two values associated to two 
topologically inequivalent phases, 
see Fig.~\ref{fig.5} (left column). 

We now disorder
the frame by shifting all site positions by random displacements of maximal amplitude $|\delta \bm x|$ while preserving the magnitude of the interactions in the corresponding 
Hamiltonian $\mathcal H_{\rm D}$. 
For sufficiently large displacements, it is impossible to keep track of the original periodic lattice, see Fig.~\ref{fig.5} (first row). 
Nonetheless, we clearly see in the third row of Fig.~\ref{fig.5} that the topologically protected zero modes located in $\mathcal B$ are preserved, despite the lack of crystalline symmetry and the impossibility to define a Bloch Hamiltonian and its 
topological 
winding numbers. Note that unlike in~\cite{Agarwala2020} both the bulk and the boundary region are 
homogeneously
disordered.
Again, the existence 
and location of a 
line of zero modes
is revealed by variations of the chiral polarization field. The variations of the orientation of $\bm \Pi(\bm x)$ occurs  over the penetration length-scale $\ell_{\rm G}$ set by the energy gap.  The coexistence of two topologically distinct amorphous phases is signalled  by a (wider) bimodal distribution of $\Pi(\bm x)$ peaked on the same values as in the pure case, see  Fig.~\ref{fig.5} (last row). This robust phenomenology is further illustrated in Supplementary Video 1, showing the evolution of the polarization field and zero-mode location as the magnitude of disorder is increased.

This observation reflects a generic feature of chiral matter.  Randomizing the frame geometry cannot alter the energy gap provided that the graph defined by the
coupling 
terms of $\mathcal H_{\rm}$ has a fixed 
chiral
connectivity.
This 
observation implies that the concept of topological phase 
{naturally applies to} 
amorphous frames that can be continuously deformed into periodic lattices. In fact, the coexistence of different  chiral 
insulators is effectively probed by the 
spatial 
distribution of the polarization field
$\Pi(\bm x)$. 
Each peak of the distribution  signals topologically inequivalent regions in amorphous chiral matter. The 
phase boundaries are then readily detected
by jumps of the chiral-polarization vector field over $\ell_{\rm G}$.\\

\noindent
{\bf{Topological zero modes of disordered chiral Hamiltonians.}}
\begin{figure*}[ht]
\begin{center}
	\includegraphics[width=\textwidth]{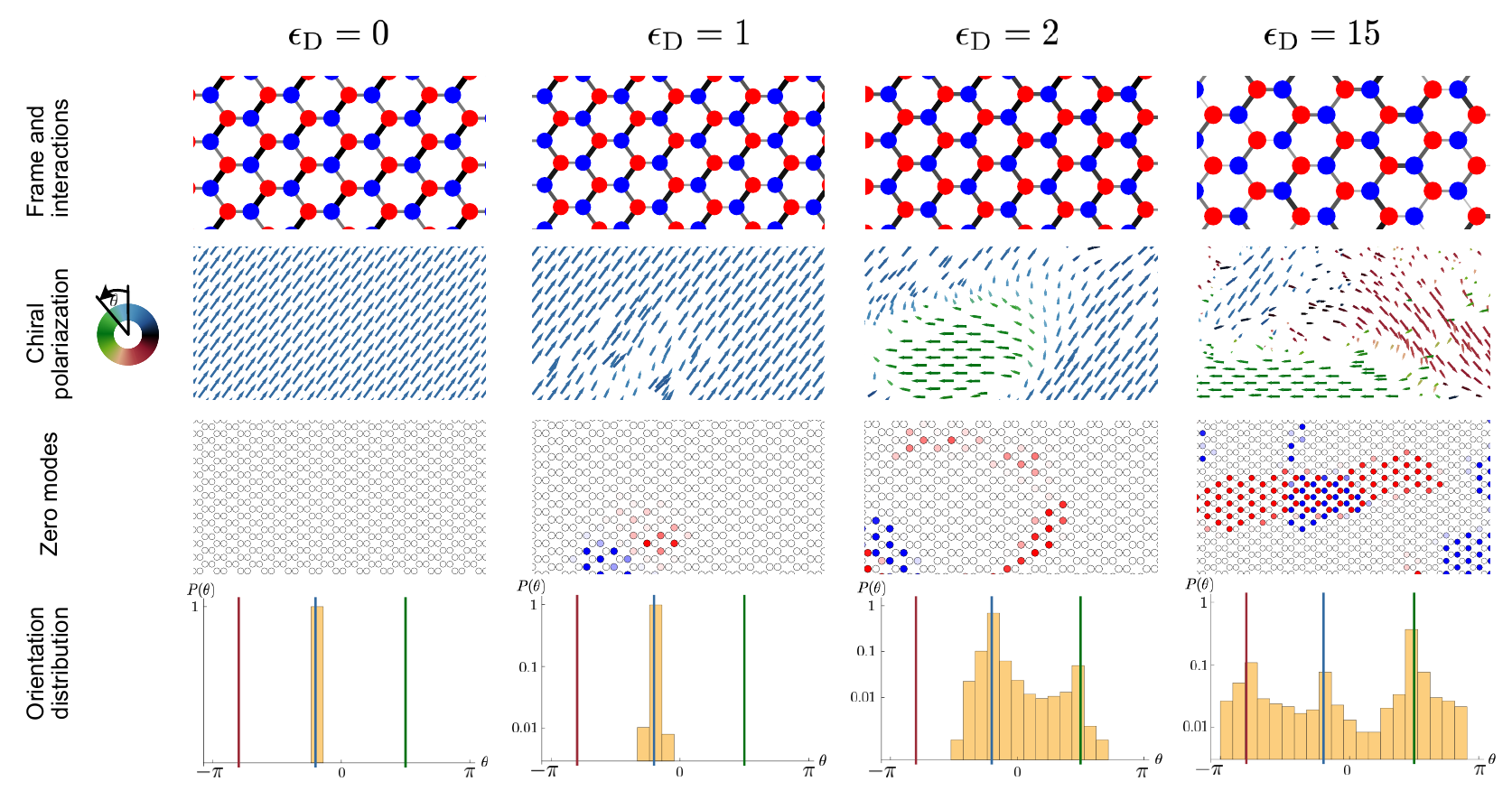}
\end{center}
\caption{
{\bf Topological zero-energy states in the bulk of disordered chiral insulators}
{First row} Sketch of the honeycomb frame and of the coupling strengths
for increasing spectral disorder. The strengths of the couplings are represented by the width of the dark lines. Their randomness is quantified by the variance of the Gaussian couplings $\epsilon_{\rm D}$. The correlation length for all the examples is $\xi=12a$. 
{Second row} Corresponding  chiral polarization fields.  The color indicates the orientation $\theta$ of $\bm \Pi(\bm x)$. 
Third Row: Magnitude of the zero-energy modes on the $A$ (red) and $B$ (blue) sites.
Fourth row: Probability density function of the orientation $\theta$. Remarkably, even in the disordered cases, the distribution peaks only at values characteristic of the three phases of the homogeneous chiral Hamiltonian.  
}
\label{fig.6}
\end{figure*}
 The case of spectral disorder is more subtle as it can trigger topological transitions. Again, we start with a concrete example. We use the same  model of insulator as in the previous section.  Considering the even simpler case of a perfect monocrystal, there is no zero mode in the sample. Keeping the frame unchanged we add disorder to the interactions in the form of random perturbations to the coupling parameters. We note $\epsilon_{\rm D}$ the width of the Gaussian disorder distribution, $\xi$ its correlation length  and $\Delta E$ the energy gap in the pure case.  
 When $\epsilon_{\rm D}/\Delta E-\ll1$ 
 no zero mode exists in this finite system see Fig.~\ref{fig.6} first column. Consistently, the local chiral polarization hardly fluctuates in space and its distribution remains peaked around the same constant value. 
 
 By contrast as $\epsilon_{\rm D}/\Delta E\sim 1$, zero energy modes emerge in the bulk. Their presence signals local 
the emergence of topologically inequivalent regions
in the material
 triggered by local gap inversions. 
The distinct phases are   revealed by the orientational  
distribution of $\bm \Pi(\bm x)$: as  disorder increases
additional peaks
grow  at values of $\theta$ characteristic of the other two  
homogeneous 
topological insulators, Fig.~\ref{fig.6} (last row). In the limit of strong disorder, the spatial extent of the coexisting phases is set by the disorder correlation length $\xi$ as exemplified in Supplementary Movie 2.  Gap closings also have  a local signature in the polarization field. As $\bm \Pi(\bm x_m)$ is only defined at the generalized Wannier centers 
(Eq.~\eqref{eq.chiralPolGeneralText}), $\bm \Pi(\bm x_m)$ cannot be computed at the center of a zero mode, which by definition does not support any Wannier mode. The proliferation of zero modes in the bulk is  therefore signaled by an increasing number of holes in the polarization field. 

The above observations do not rely on the specific model we use in Figs.~\ref{fig.5} and~\ref{fig.6}. Generically, adding spectral disorder to a chiral Hamiltonian results in the nucleation of additional topological phases decorated by zero modes at their boundaries. Even in the absence of a Bloch theory, we can distinguish the topological nature of the coexisting phases by measuring their average chiral polarization. For spatially correlated disorder the spatial extent of each phase is set by the disorder correlation length $\xi$.\\

\begin{figure*}[ht]
\begin{center}
	\includegraphics[width=\textwidth]{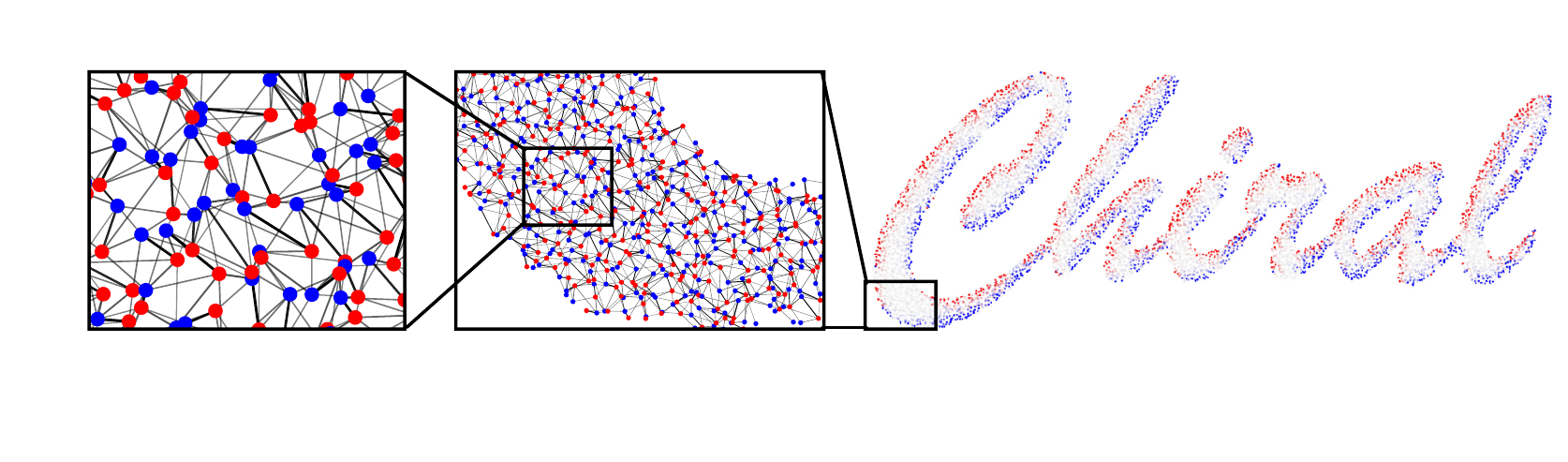}
\end{center}
\caption{
{\bf Disordered chiral metamaterial}
Macroscopic view and close ups on an amorphous frame supporting a disordered chiral insulator. The frame is defined adding a strong positional disorder to a Honeycomb lattice $|\delta \bm x|=a$. Using the same Hamiltonian as in Figs.\ref{fig.5} and~\ref{fig.6}, we add spectral disorder corresponds to $\epsilon_{\rm D}=2$. Cutting the sample to form the word ''chiral'' reveals a continuous distribution of zero modes along the edge.
\label{fig.7}
}
\end{figure*}

\noindent
{\bf{Designing  topologically protected zero modes in amorphous chiral matter.}}
It is worth stressing that disordered chiral insulators generically support topologically protected zero-energy modes at their boundaries.  Unlike crystaline topological insulators,  the lower the bulk and edge symmetries the more robust the edge states. 

Cutting   an amorphous sample into two parts without inducing the proliferation of boundary zero modes is virtually impossible. It would require cutting bonds while preserving the connectivity between all pairs of $A$ and $B$ site connected by the local polarization vectors $\bm \Pi(\bm x_m)$; only this type of configurations can be continuously deformed into crystals having edges matching that  of tilings generated by the unit cell of an atomic limit. These cuts require extreme fine tuning in macroscopic samples and are therefore virtually impossible to achieve.
This property makes the design of zero energy wave guides very robust in amorphous chiral matter. As illustrated in Fig.~\ref{fig.7}.\\

\david{}{
\noindent
{\bf{Measuring the chiral polarization.}}
In this section we show that the chiral polarization is not only a powerful theoretical concept, but an actual material property readily accessible to experiments. Two scenarios are possible: when the (low energy) eigenfunctions can be measured, 
the chiral polarization can be directly evaluated using its definition, Eq.~\eqref{eq:defPolarization}. This 
technique is straightforward e.g. in mechanical metamaterials~\cite{serra2018observation}, where the vibrational eigenmodes can be imaged in real space in response to mechanical actuation.

Alternatively, when spectral properties are out of reach of quantitative measurements, we can infer the value of  the chiral polarization from the dynamic spreading of localized chiral excitations. This approach builds and generalizes the technique  pioneered in the context of periodically driven photonic quantum walk~\cite{cardano2017detection,maffei2018topological}. For the sake of clarity we henceforth limit our discussion to 1D, two-band insulators although the reasonning applies in higher dimensions.
We introduce the dynamical chiral polarization
$\Pi_\Psi (t)= \expval{\amsmathbb{C}  \hat X }{\Psi(t)}$ 
defined over the  time-evolved states $\Psi(t) = \exp(-i H t) \Psi(0)$, where $\Psi(0)$ is a localized chiral state. Should one be able to initalize an experiment in a Wannier State $\Psi(0) = W_{n,\bm R}$, the wave function would spread as in Fig.~\ref{fig.9}a, but remarkably the dynamical chiral polarization $\Pi_\Psi (t)$ 
would be stationnary and equal to $\Pi$
 in a homogeneous system as illustrated in Fig.~\ref{fig.9}a, and demonstrated in the Method section. In practice, it would be always easier to approximate the Wannier state by  excitations $\Psi_{AB}$ (resp. $ \Psi_{BA}$)
 localized on two neighboring $A$ and $B$ sites (resp. $B$ and $A$). The result of this procedure is shown in Fig.~\ref{fig.9}b and reveals that the long-time dynamics of $\Pi_\Psi (t)$ converges towards the  chiral polarization $\Pi$. However, we stress that the essential information about the orientation of $\Pi$ is already accessible at very short times and would not suffer from possible damping issues. When $\Pi_\Psi (t=0)$ and $\Pi$ have opposite signs, we observe very large amplitude oscillations reflecting the dynamic reversal of the chirality of the wave packet at short times. Conversely when $\Pi_\Psi (t=0)$ and $\Pi$ are parallel the convergence is very fast and devoid of large amplitude fluctuations. 

It is worth noting that the chiral initial state $\Psi(t=0) = \Psi_{AB}$ is an atomic-limit eigenstate. The dynamics can then be seen as the result of a quench at $t=0$ starting from the atomic-limit Hamiltonian.
The  amplitude of the fluctuations in Fig.~\ref{fig.9}b then reveals the topological nature of the quench. 
As a last comment we stress that although our protocol is close to the chiral displacement method introduced and used in~\cite{cardano2017detection,maffei2018topological,st2020measuring,d2020bulk}, it is not tight to a Bloch Hamiltonian model, but characterizes an intrinsic (meta)material property.

\begin{figure*}
\begin{center}
	\includegraphics[width=17cm]{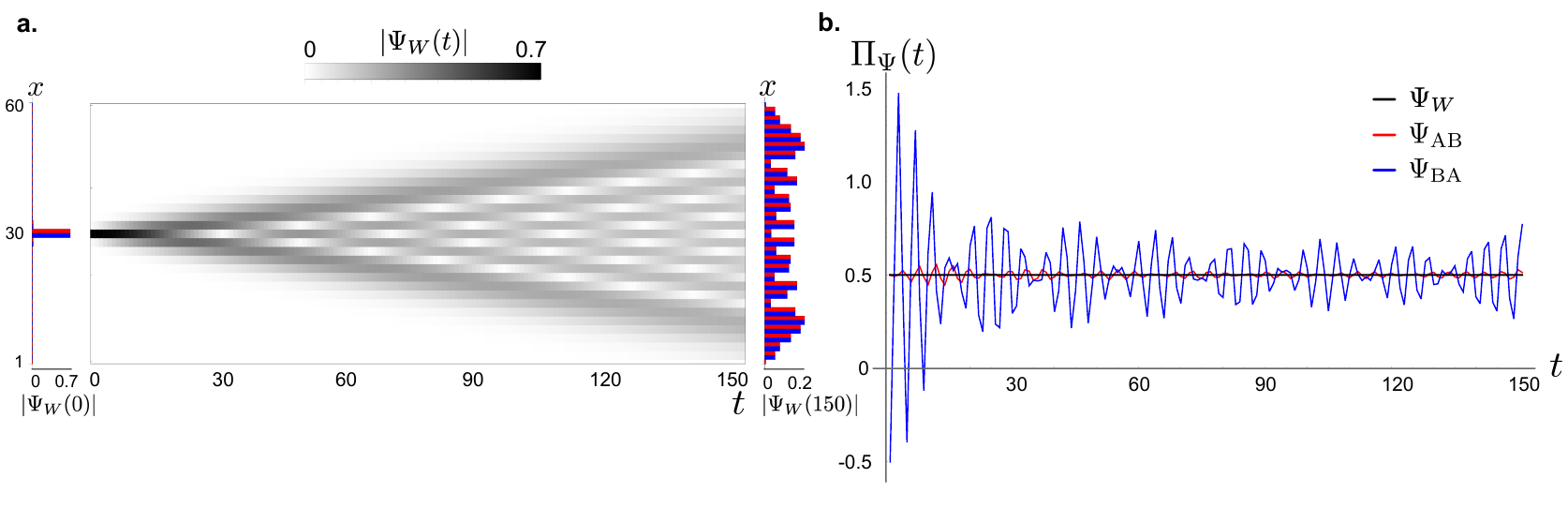}
\end{center}
\caption{
\david{}{
{\bf Measuring the chiral polarization in time.}
{\bf a.} Left: 
Dynamical evolution of a Wannier state  in the ground state of a two-band 
SSH model, with hopping ratio $t_1/t_2=0.1$. The state is localized 
in the middle of a finite system of 60 unit cells. 
Center: Time evolution of the wave-function amplitude.
Right: The amplitude of the 
final state at time $t=250$ is represented 
on the A (red) and B (blue) sites of the lattice. 
{\bf b.} The dynamical chiral polarization 
$\Pi_{\psi}(t)=
\expval{\amsmathbb C \hat X }{\Psi (t)}$  corresponding to the protocol described in {\bf a} 
 is constant in time 
(black solid line). 
By comparison, the dynamical chiral polarization 
starting from a state 
$\Psi_{AB}(t=0)$ (resp. $\Psi_{BA}(t=0)$), 
localized  on two neighboring sites 
$A$ and $B$  (resp. $B$ and $A$)
shows fluctuations around the static chiral polarization whose amplitude depends on the initial state. 
The sign  $\Pi_\psi(t)$ is reversed at short time when the chiral polarization of the  initial state is opposite to the static chiral polarization of the SSH chain. This results in large amplitude  oscillations. The short time dynamics of $\Pi_\psi$ therefore provides a direct access to the orientation of the material chiral polarization.
}}
\label{fig.9}
\end{figure*}

}

\section*{Discussion}

We have established a generic framework  to characterize, elucidate and design the topological phases of chiral insulators. In crystals, we  show that the frame topology and the frame geometry conspire with  Bloch Hamiltonian topology to determine the zero-mode content of the bulk and interfaces. In the bulk, the frame topology fully determines the algebraic number of zero-energy modes counted by the chiral charge $\mathcal M$.  Chiral insulators, however, are distinguished  one another via their chiral polarization  $\bm \Pi$ set both by the  frame geometry and Bloch-Hamiltonian topology. At their surface, the number of zero-energy states is prescribed by the  interplay between the Bloch Hamiltonian topology and the  frame  geometry  in the bulk on one hand, and by the frame topology of the boundary on the other hand. This subtle tango goes beyond  the bulk-boundary-correspondence principles solely based on Hamiltonian topology. %

We have shown that chiral symmetry alone translates real-space properties into spectral phases without relying on any crystalline symmetry and translational invariance when expressed as a sublattice symmetry. Chiral symmetry does not merely complement the classification of topological quantum chemistry~\cite{Kruthoff2017,Bradlyn2017,Tang2019,Zhang2019} but also makes it possible to distinguish topological phases in amorphous matter. In disordered system, introducing the concept of chiral polarization field, we  provide a practical platform to  detect topological phases coexisting in disordered samples, an to design robust zero-mode wave  guides at their boundaries.

We expect our framework to extend beyond Hamiltonian dynamics when dissipative processes obey the chiral symmetry~\cite{Kawabata2019}. We therefore conjecture that real-space topology, geometry and non-Hermitian operator topology should cooperate in chiral dissipative materials as diverse as cold atoms to photonics, robotic devices and active matter.

\begin{acknowledgements}

We acknowledge support from ANR WTF, and ToRe IdexLyon breakthrough programs. We also thank  J. Asboth, A. Bernevig, A. Dauphin, K. Gawedzki, A. Grushin,  Y. Hatsugai,  P. Massignan, A. Po, A. Schnyder and A. Vishwanath for insightful discussions. 

\end{acknowledgements}

\appendix

\section*{Methods}
\label{eq:BlochTheory}
\noindent
{\bf{Conventions for the Bloch decomposition.}}
For the sake of clarity, we first  introduce the main quantities used throughout all the manuscript to describe waves in periodic lattices. We note $\ket{\Psi_{n,\bm k}}$ the Bloch eigenstates. They correspond to wavefunctions $\bra{\bm x}\ket{\Psi_{n,\bm k}}=\varphi_{n,\bm k}(\bm x)e^{i\bm k\cdot\bm x}$, where  ${\bm k}$ is the momentum in 
the Brillouin Zone (BZ), and where the normalized function $\varphi_{n,\bm k}$ has a periodicity of one unit cell~\cite{Vanderbilt}. In this article, we use the following convention to express the Bloch states as a superposition of plane waves:
\begin{equation}
\ket{\Psi_{n,\bm k}}=\sum_{\alpha}u_{n,\alpha}(\bm k)\ket{\bm k,\alpha},
\label{eq:BlochConvention}
\end{equation}
 where $\alpha$ labels the different atoms in the crystal, and $\ket{\bm k,\alpha}$ represents the Fourier transform of the real-space position basis: $\ket{\bm k,\alpha}=\sum_{\bm R} \exp(i\bm k\cdot \bm R)\ket{\bm R+\bm r{_\alpha}}$, $\bm R$ being a Bravais lattice vector and $\bm r_{\alpha}$ a site position within the unit cell. 
 We stress that here the components  $u_{n,\alpha}(\bm k)$ are  periodic functions of $\bm k$ over the BZ. It is worth noting, however, that there exists multiple conventions to decompose the Bloch states as discussed e.g in the context of graphene-like systems in~\cite{Bena2009,Fuchs2010,Fruchart2014}. A common alternative uses nonperiodic components over the BZ which carry an additional phase encoding the position of each atom within the unit cell:   $\ket{\Psi_{n,\bm k}}=\sum_{\alpha}\tilde u_{n,\bm k,\alpha}e^{i\bm k\cdot \bm r_{\alpha}}\ket{\bm k,\alpha}$. We will comment on the translation of our results from one convention to the other in the following.%
 
\noindent
{\bf{Wannier functions.}}
By definition the Wannier function associated to a Bloch eigenstate is given by the inverse Fourier transform (up to a phase):
\begin{equation}
\ket{W_{n,\bm R}}=\int_{\bm k}\!e^{-i\bm k\cdot\bm R}\ket{\Psi_{n,\bm k}},
\label{eq:defWannier}
\end{equation}
where  
$\int_{\bm k}\cdot\equiv \Omega^{-1} \int_{\BZ}\! {\rm d}^d{k}\,\cdot$,
 $\Omega$ being the volume of the BZ.
Note that for sake of clarity, we here and henceforth assume that the spectrum does not include band crossings. The technical generalization of our results to degenerated spectra is straightforward but involves some rather heavy algebra, see e.g.~\cite{Vanderbilt}.\\

\noindent
{\bf{Projected position operator and sublattice Zak phases.}} Ignoring the distinction between the $A$ and $B$ sites, we can first compute the action of the position operator on the Wannier states following ~\cite{Vanderbilt}:
\begin{align}
\bra{\bm x}\widehat X \ket{W_{n,\bm R}}&=\int_{\bm k}\!\bm x \,e^{i\bm k \cdot(\bm x-\bm R)}{\varphi_{n,\bm k}}(\bm x)\nonumber\\
&=\int_{\bm k}\!\left(-i\partial_{\bm k}e^{i\bm k\cdot(\bm x-\bm R)}+\bm Re^{i\bm k\cdot(\bm x-\bm R)}\right){\varphi_{n,\bm k}}(\bm x)\nonumber\\
&=\int_{\bm k}\!e^{-i\bm k\cdot \bm R}\left[e^{i\bm k\cdot \bm x}\left(\bm R+i\partial_{\bm k}\right)\right]{\varphi_{n,\bm k}}(\bm x),
\label{eq:xW}
\end{align}
where in the last step we applied an integration by parts, using that 
$\ket{\Psi_{n,\bm k}}= \ket{\Psi_{n,\bm k+\bm G}}$
with $\bm G$ a primitive reciprocal vector. 
The generalization of Eq.~\eqref{eq:xW} to the position operator projected on  the  sublattice  $a=A,B$ is straightforward:
\beq
\bra{\bm x}\widehat X\amsmathbb{P}^a\ket{W_{n,\bm R}}=
\int_{\bm k}\!e^{-i\bm k\cdot \bm R}\left[e^{i\bm k\cdot \bm x}
\left(\bm R+i\partial_{\bm k}\right)\right]\amsmathbb{P}^a{\varphi_{n,\bm k}}(\bm x),
\eeq
which allows us to define the average positions $\expval{\bm x ^a}_{n,\bm R}$ restricted to the site  $a=A,B$ and to the $n^{\rm th}$ band excitations:
\begin{align}
\expval{\bm x ^a}_{n,\bm R}\equiv 
&\expval{\amsmathbb{P}^a \hat X\amsmathbb{P}^a}{W_{n,\bm R}}\nonumber\\
=&{\bm R}\int_{\bm k}\! \expval{\amsmathbb{P}^a}{\varphi_{n,\bm k}}
+\frac{1}{\Omega}\bm{\Gamma}^a_{\text{Zak}}(n),
\label{eq.chiralX1}
\end{align}
where $\Omega$ is the volume of the BZ, $\ket{\varphi_{n,\bm k}}=e^{-i\bm k\cdot \hat X}\ket{\Psi_{n,\bm k}}$, and $\bm{\Gamma}^a_{\text{Zak}}(n)$ is the vector composed of the $d$ sublattice Zak phases associated to the $n$-th band: 
\beq
\bm{\Gamma}^a_{\text{Zak}}(n)=
i\Omega\int_{\bm k}\!\expval{\amsmathbb{P}^a \partial_{\bm k}\amsmathbb P^a}{\varphi_{n,\bm k}}.
\label{eq:defZakn}
\eeq
We can further simplify Eq.~\eqref{eq.chiralX1} noting that  the orthonormality of the $\ket{\varphi_{n,\bm k}}$ implies $\expval{\amsmathbb{P}^A+\amsmathbb{P}^B}{\varphi_{n,\bm k}}=1$ and $\expval{\amsmathbb{P}^A-\amsmathbb{P}^B}{\varphi_{n,\bm k}}=0$, which yields $\expval{\amsmathbb{P}^a}{\varphi_{n,\bm k}}=1/2$. All in all, we find a simple relation between the average of the position operator and the Zak phase of the Bloch eigenstates over the BZ:   
\beq
\expval{\bm x ^a}_{n,\bm R}=\frac{\bm R}{2}+\frac{1}{\Omega}\bm{\Gamma}^a_{\text{Zak}}(n).
\label{eq.chiralX2}
\eeq

\noindent
{\bf{Chiral polarization and sublattice Zak phases.}}
We are now equipped to compute the chiral polarization, defined as the difference between the expected value of the projected position operators over the occupied eigenstates ($n<0$).  It readily follows from Eq.~\eqref{eq.chiralX2} that $\bm \Pi$ corresponds to the difference of the sublattice Zak phases:
\begin{align}
\bm{\Pi}\equiv& 2 \sum_{n<0}\expval{\bm x ^A}_{n,\bm R}-\expval{\bm x ^B}_{n,\bm R}\nonumber\\
=&\frac{2}{\Omega}\sum_{n<0}\bm{\Gamma}^A_{\text{Zak}}(n)-\bm{\Gamma}^B_{\text{Zak}}(n).
\label{eq:PiZak}
\end{align}
Two comments are in order. Firstly,  the sum could have been also taken over the unoccupied states ($n>0$). 
As $\amsmathbb{C}^2=\amsmathbb{I}$, the sublattice phase picked up by $\ket{\varphi_{n,\bm k}}$ is indeed the same as that of its chiral partner $\ket{\varphi_{-n,\bm k}}=\amsmathbb C \ket{\varphi_{n,\bm k}}$.
Secondly, we stress that Eq.~\eqref{eq:PiZak} does not depend on the specific convention of the Bloch representation. This relation, however does not disentangle   the respective contributions of the frame geometry and of the Hamiltonian on the chiral polarization. 
To single out the two contributions, we now use the  specific Bloch representation 
\eqref{eq:BlochConvention}. 
Given this choice, the sublattice Zak phase is naturally divided into two contributions leading to
\begin{align}
\bm{\Gamma}^a_{\text{Zak}}(n)=&\Omega\int_{\bm k}\,\sum _{\alpha \in a}\left(u^*_{n,\alpha}u_{n,\alpha}\bm r_{\alpha}+iu^*_{n,\alpha}\partial_{\bm k}u_{n,\alpha}\right).
\label{eq:Zakn}
\end{align}
The first term on the r.h.s. is  the intracellular contribution to the Zak phase  while the second is proportional to the sublattice intercellular Zak phase following 
to the definitions of ~\cite{Rhim:2017}
\begin{equation}
\gamma_j^a(n)\equiv i \int \!{\rm d}k_j\sum_{\alpha \in a}u^*_{n,\alpha}(\bm k)\partial_{k_j}u_{n,\alpha}(\bm k).
\end{equation}
Summing Eq.\eqref{eq:Zakn} over all occupied bands, and using the orthogonality of the chiral component $u_{n,\alpha}$ we then recover our central result:
\beq
 \bm{\Pi}=
  \bm p +
  \frac{2}{\Omega^{1/d}}\left(\bm{\gamma}^A-\bm{\gamma}^B\right),
\label{eq:Pi_Main}  
\eeq
 where $\bm p=\sum_{\alpha\in A} \bm r_{\alpha} -\sum_{\alpha\in B} \bm r_{\alpha} $ is the geometrical polarization of the corresponding unit-cell and 
 $\bm{\gamma}^a=\sum_{n<0}\bm{\gamma}^a(n)$. The chiral polarization is the sum of one contribution
 coming only from the frame geometry and one contribution characterizating the geometrical phase of
 the Bloch eigenstates.\\

\noindent
{\bf{Chiral polarization in different Bloch conventions.}}
Although the physical content of the chiral polarization does not depend on the choice of the Bloch convention, it is worth explaining how to derive its functional form for the other usual representation where $\ket{\Psi_{n,\bm k}}=\sum_{\alpha}\tilde u_{n,\alpha}(\bm k)e^{i\bm k\cdot \bm r_{\alpha}}\ket{\bm k,\alpha}$. Within this convention the total Zak phase takes the form 
\beq
\bm{\Gamma}_{\text{Zak}}^a(n)=i\int_{\BZ}\! \! \! \! {\rm d}^d{k}\,\sum_{\alpha\in a}\tilde u^*_{n,\alpha}\partial_{\bm k}\tilde u_{n,\alpha},
\eeq
which does not allow the distinction  between the geometrical and the Hamiltonian contributions to $\Pi$ when performing the sum over the occupied band in  Eq.~\eqref{eq:PiZak}. This observation further justifies our choice for the Bloch representation.\\

\noindent{\bf Quantization of the intercellular Zak-phase in chiral insulators.}
\label{Appendix:Zak}
To demonstrate the quantization of $\gamma_j=\gamma_j^A+\gamma_j^B$, we resort to the Wilson loop formalism reviewed e.g. in Ref.~\cite{Neupert:2018}.

Let us first recall the definition of the non-Abelian Berry-Wilczek-Zee connection along the Brillouin zone for a 
set of smooth vectors $\ket{u_n({\bm k})}, n=1,...M$: 
\begin{equation}
    \mathbf{A}_{nm}({\bm k}) = 
    \bra{u_n ({\bm k}) }\partial_{\bm k} \ket{u_m({\bm k})}.
\label{eq:BerryNonAb}
\end{equation}
The associated Wilson loop operator defined along the path $\mathcal C_j$ through the Brillouin zone is given by the ordered exponential 
\begin{equation}
    W_j = \overline{\textrm{exp}} 
    \left( - \int_{\mathcal C_j} d{\bm k}\cdot \mathbf{A}({\bm k}).
    \right)
    \label{eq:Wilson}
\end{equation}
The topological properties of a generic gapped chiral Hamiltonian are conveniently captured by  smooth deformations yielding a flat spectrum $E = \pm 1$. The corresponding Bloch Hamiltonian is then given by
\begin{equation}
    H = 
        \begin{pmatrix}
        0 & Q({\bm k}) \\
        Q^\dagger ({\bm k}) & 0 
        \end{pmatrix}
\label{eq:FlattenH}
\end{equation}
where $Q({\bm k})$ is a nonsingular unitary matrix. Without loss of generality, we write the corresponding eigenstates as
\begin{equation}
    \ket{u_{\pm n} ({\bm k})} = 
    \frac{1}{\sqrt{2}}
        \begin{pmatrix}
        \pm Q ({\bm k}) \ket{e_n^{B}} \\
        \ket{e_n^{B}}
        \end{pmatrix}
\end{equation}
where the sign $\pm $ identifies the sign of the eigenvalue
$E=\pm 1$ and the normalized vectors $\ket{e_n^{B}}$ form a basis of the Hilbert
space of $Q^\dagger$.
The non-Abelian connection \eqref{eq:BerryNonAb} for the negative (resp. positive) energy states then takes the simple form 
\begin{align}
    \mathbf{A}^{-}_{nm}({\bm k}) & = 
    \frac12 
    \bra{e_n^{B}} Q^\dagger({\bm k}) \partial_{\bm k} Q({\bm k}) \ket{e_m^{B}}
\label{eq:BerryConnection-Winding}
\\
    &= 
    \mathbf{A}^{+}_{nm}({\bm k}) 
\end{align}
It follows from the definition of the Wilson-loop operator (Eq.~\eqref{eq:Wilson}) that the intercellular Zak phase for the negative energy bands $\gamma=\gamma^A+\gamma^B$ is defined in terms of the Wilson loops for the non-Abelian connection $\mathbf{A}^{-}({\bm k}) $ as 
\begin{equation}
    \gamma_j  = 
    -i\ln \det W_j^{-}
\end{equation}
The quantization of all $d$ intercellular Zakk phases then  follows from  Eqs~\eqref{eq:Wilson} and \eqref{eq:BerryConnection-Winding}: 
\begin{align}
    \gamma_j &  =  -i \tr \ln \left [\overline{\textrm{exp}} 
    \left( - \frac12 \int_{\mathcal C_j} d{\bm k}\cdot 
         \partial_{\bm k} \ln Q({\bm k})\right)\right]
     \\
     &=  \pi w_j \textrm{ mod }(2\pi)
     \label{eq:gammaw}
\end{align}
where the $\mod (2\pi)$ indetermination stems from the choice of the  branch cut of the complex $\ln$ function, and where $w_j$ is the standard winding of the chiral Hamiltonian \eqref{eq:FlattenH}: 
\begin{align}
w_j &=\frac{i}{4\pi}\int_{\mathcal C_j}{\rm d}{\bm k}\cdot \tr\left[\partial_{{\bm k}}H\amsmathbb{C}H^{-1}\right]\in\amsmathbb Z,
     \label{eq:wjappendix}
    \\
    &=\frac{1}{2\pi i}\int_{\mathcal     C_j}{\rm d}{\bm k}\cdot
    \tr\left[Q^{-1} \partial_{\bm k}Q \right].
\end{align}
We therefore conclude that  the $d$ Zak phases are topological phases defined modulo $2\pi$.  \\

\noindent
{\bf{Relating the sublattice Zak phases to the winding of the Bloch Hamiltonian.}}
\label{Appendix:Zak_Winding}
We here demonstrate the essential relation given by Eq.~\eqref{eq:defw}. To do so, we  relate the winding $w_j$ to the sublattice Zak phases by evaluating  the trace in Eq.~\eqref{eq:wjappendix} using the eigenstate basis. Noting that $	\expval{\partial_{\bm k}H({\bm k})\amsmathbb{C}H^{-1}({\bm k})}{ u_{n}}
	=
	- 2 \expval{\amsmathbb{C} \partial_{\bm k} }{ u_{n}}$, the winding takes the simple form
\begin{align}
w_j &=-\frac{i}{2\pi}\int_{\mathcal{C}_j} {\rm d}{ k} \sum _n \expval{\amsmathbb{C}\partial_{ k}}{ u_{n}}.
\end{align}
Decomposing the chiral operator on the two sublattice projectors $\amsmathbb C=\amsmathbb P^A-\amsmathbb P^B$, yields 
\begin{equation}
\pi w_j =\left(\gamma_j^B-\gamma_j^A\right)\in \pi \amsmathbb Z.
\label{eq:wgamma}
\end{equation}

\noindent
{\bf{Quantization of the  sublattice Zak phases.}}
Eqs.~\eqref{eq:gammaw} and ~\eqref{eq:wgamma} shows that both the sum and the difference of the sublattice Zak phases are quantized:
\begin{align}
\gamma_j^A+\gamma_j^B&=\pi w_j+2\pi m,\;\;\;\;\;m\in\pi\amsmathbb{Z},\nonumber\\
\gamma_j^B-\gamma_j^A&=\pi w_j.
\end{align}
It then follows that both sublattice phases $\gamma_j^A$ and $\gamma_j^B$ are integer multiples of $\pi$.\\

\noindent
{\bf{How does the winding number of a chiral Bloch Hamiltonian change upon unit cell redefinition?}}
\label{Appendix.TransformationU}
Starting from a chiral Hamiltonian $\mathcal H$, we demonstrate below the relation between the winding numbers associated  to the  Bloch Hamiltonians constructed from  different choices of unit cells,  Eq.~\eqref{eq:indexP}.

The definition of Bloch waves and Bloch Hamiltonians require  prescribing  a unit cell. Starting with a first choice of a unit cell geometry, say unit cell (1), we can write $H^{(1)}(\bm k)$ in the chiral basis as
\beq
H^{(1)}({\bm k})=\begin{pmatrix}0&Q^{(1)}\\{Q^{\dag}}^{(1)}&0\end{pmatrix},
\eeq
Let us now opt for a second choice of unit cell, say choice (2). The Bloch Hamiltonians $H^{(1)}$ and  $H^{(2)}$ are then related by a unitary transformation
\begin{equation}
H^{(2)}=U^\dagger H^{(1)}U,
\label{eq:Udef}
\end{equation}
where  the components of the unitary matrix are given by
\begin{equation}
U_{\alpha\beta}=\exp(i{\bm k}\cdot \bm R_\alpha^{(12)})\delta_{\alpha\beta},
\label{eq:defuappendix}
\end{equation}
where the $\bm R^{12}_\alpha$ are the Bravais vectors connecting the position of the atoms in the two unit-cell conventions, see Fig.~\ref{fig.Raplha} for a simple illustration. We note that, we have implicitly  ignored  the trivial redefinitions of the unit cell that reduce to permutations of the site indices. 
\begin{figure}
\begin{center}
	\includegraphics[width=\columnwidth]{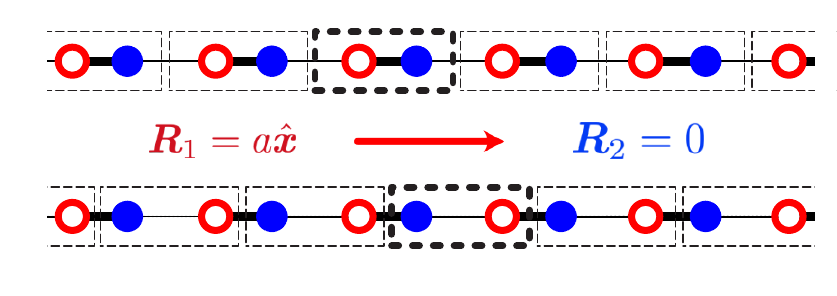}
\end{center}
\caption{{\bf Unit cell transformation.} We illustrate the definition of the $\mathbf R_\alpha$ vectors using the simple example of a SSH chain. For the first atom (empty symbol) $\mathbf R_1=a\hat{\mathbf x}$ while $\mathbf R_2=0$ for the second atom (solid symbol).
}
\label{fig.Raplha}
\end{figure}
We can then express the winding of $H^{(2)}$ using Eq.~\eqref{eq:Udef} in the definition of Eq.~\eqref{eq:wjappendix}, which yields
\begin{equation}
w_j^{(2)} =\frac{i}{4\pi}\int_{\mathcal{C}_j} \!\!d{\bm k} 
	\tr \left[\partial_{\bm k} (U H^{(1)} U^{\dagger})\amsmathbb{C} (U H^{(1)}U^{\dagger})^{-1}   \right].
\end{equation}
Expanding the gradient, using the trace cyclic property and noting that $\left[\amsmathbb{C},U\right]=0$, we  find
\begin{align}
w_j^{(2)} =w_j^{(1)}-\frac{i}{2\pi}\int_{\mathcal C_j}d{\bm k}\tr\left[\partial_{{\bm k}}U\amsmathbb C U^{-1}\right].
\end{align}
This equation relates the winding numbers of the two Bloch Hamiltonians to the winding number of the transformation matrix $U$, which is by definition a  geometrical quantity independent of $\mathcal H$.  Using Eq.~\eqref{eq:defuappendix} leads to the remarkable relation which relates the spectral properties of the Hamiltonian to  the unit-cell  geometry
\begin{align}
w_j^{(1)}-w_j^{(2)}&=\frac{i}{2\pi}\int_{\mathcal C_j}d{\bm k}\tr\left[\partial_{{\bm k}}U\amsmathbb C U^{-1}\right]\nonumber\\
&=\frac{1}{a_j}\left(\sum_{\alpha\in A}R_{\alpha}-\sum_{\alpha\in B}R_{\alpha}\right).
\end{align}
\medskip

\noindent
{\bf Zero energy flat-band insulators.}
\david{}{
We consider a flat-band chiral insulator, defined on a lattice with an non-vanishing chiral
charge. In mechanics this situation is readily achieved adding extra bonds to further rigidify an otherwise isostatic lattice.  It is characterized by a finite gap separating positive and negative energy states and by an additional flat band at $E=0$. 
In such a phase,  there may exist additional zero energy edge states in addition to the bulk zero-energy modes. These edge states are analogous to to the topological edge modes of insulators. Our goal is here to  derive a bulk-boundary correspondence for these materials and provide a count of their zero-energy edge states. We will show that this correspondence involves the specific geometry of the eigenstates as opposed to their topology in the case of genuine insulators. 

To show this we  first derive the expression of the
 chiral polarization in the presence of a finite bulk chiral charge. 
Our starting point is Eq.~\eqref{eq:PiZak}, which relates to the chiral polarization of a crystal to the sublattice Zak phases given by Eq.~\eqref{eq:defZakn}:
\begin{align}
\bm{\Pi}\equiv& 2 \sum_{n<0}\expval{\bm x ^A}_{n,\bm R}-\expval{\bm x ^B}_{n,\bm R}\nonumber\\
=&\frac{2}{\Omega}\sum_{n<0}\bm{\Gamma}^A_{\text{Zak}}(n)-\bm{\Gamma}^B_{\text{Zak}}(n).
\label{eq:Pi_FB}
\end{align}
The sum over all the negative energy bands $n<0$ is half the sum over the non-zero energy states 
$n\neq 0$ given by
\begin{align}
    \sum_{n\neq0}\bm\Gamma^{a}(n)&=\Omega\int_{\bm k}\sum_{\alpha\in a}\sum_{n}u_{n,\alpha}^*u_{n,\alpha}\bm r_{\alpha}+\frac{2}{\Omega^{1/d}}\bm\gamma^a\nonumber\\
    &=\Omega\int_{\bm k}\sum_{\alpha\in a}\left(1-\sum_{n_0}u_{n_0,\alpha}^*u_{n_0,\alpha}\right)\bm r_{\alpha}+\frac{2}{\Omega^{1/d}}\bm\gamma^a.
\end{align}
In the last line, we single out the role of the bulk zero-energy modes
indexed by $n_0$. 
Using the above expression to evaluate the r.h.s. of Eq.~\eqref{eq:Pi_FB}, we find an expression similar
to Eq.~\eqref{eq:Pi_Main} in the main text:
\beq
 \bm{\Pi}= (\bm p -\bm p_{\text{ZM}})+
  \frac{2}{\Omega^{1/d}}\left(\bm{\gamma}^A-\bm{\gamma}^B\right).
  \label{eqSI:Pi}
\eeq
A first noticeable difference with Eq.~\eqref{eq:Pi_Main} is a spectral correction to the geometrical polarization stemming from the localized zero-energy bulk modes. This zero-mode polarization is given by
\beq
\bm p_{\text{ZM}}=-\Omega\int_{\bm k}\sum_{n_0}\left(\sum_{\alpha\in A}-\sum_{\alpha\in B}\right)u_{n_0,\alpha}^*u_{n_0,\alpha}\bm r_{\alpha}.
\eeq
Three comments are in order. Firstly, we stress that  while the geometrical polarization $\bm p$ depends on the choice of origin in the 
presence of an excess of chiral charge, the difference $\bm p-\bm p_{\text{ZM}}$, and $\Pi$, are both independent of the 
 frame's origin.  Secondly, unlike in insulators, the difference between the intercellular sublattice  Zak phases, $\bm \gamma^A-\bm\gamma^B$ is  does not
identify with the winding number of the Bloch Hamiltonian. 
In fact it is not a topological quantity: it continuously depends on the  
specific couplings of the Hamiltonian. 
Finally, we point that, by definition, the chiral polarization does not depend on the Bloch convention.
A change in the Bloch convention changes the geometrical polarization, the zero-mode polarization, and the intercellular zak phases in such a way that all  corrections cancel one another.

Equiped with Eq.~\eqref{eqSI:Pi}, we now now turn to the generalization of the bulk boundary correspondence for flat-band insulators. 
We consider a crystalline material $\mathcal S$ terminated by a clean edge $\partial\mathcal S$ oriented
along the Bravais vector $\bm a_1$. This edge may host $\mathcal V^{\text{NT}}$ non-trivial zero-energy modes, in addition to the  (trivial) bulk zero modes associated to the flat  band.
The edge defines a unit cell that may not be compatible with that of the atomic limit.%
We can nonetheless extend the edge region such that it matches the unit-cell compatible with the atomic limit (AL). 
The idea being that $\mathcal V^{\text{NT}}$ is fully determined by the additional chiral charge of the edge with respect to that provided by the bulk chiral charge density.  
Following the same reasoning as in the main text, this extra chiral charge 
is given by the difference of geometrical polarization and zero-mode polarization:
\beq
\mathcal V^{\text{NT}}=\mathcal N^{\mathcal B}
\left[(p_2-{p_{\text{ZM}}}_2)_{\text{AL}}-(p_2-{p_{\text{ZM}}}_2)\right],
\eeq
where  $\mathcal N^{\mathcal B}$ is the boundary length expressed in units of unit-cell length. The first term is computed in the unit cell compatible with the atomic limit, and the second term is computed in the original unit cell defined by the edge $\partial \mathcal S$.\\
The invariance of the chiral polarization with respect to unit cell transformations allows the connection with the intercellular sublattice Zak phase:
\begin{multline}
\left(p_2-{p_{\text{ZM}}}_2+
  \frac{2}{\Omega^{1/d}}\left(\gamma_2^A-\gamma_2^B\right)\right)_{\text{AL}}
\\
=p_2-{p_{\text{ZM}}}_2+
  \frac{2}{\Omega^{1/d}}\left(\gamma_2^A-\gamma_2^B\right),
\end{multline}
where AL denotes the terms evaluated in the unit-cell compatible with the atomic limit.
All in all, the non-trivial zero-energy content of flat band insulators 
is given by a formula whhich generalizes Eq.~\eqref{eq:KaneLubensky}:
\beq
\mathcal V^{\text{NT}}=\mathcal N^{\mathcal B}\frac{2}{\Omega^{1/d}}\left[(\gamma_2^A-\gamma_2^B)-(\gamma_2^A-\gamma_2^B)_{\text{AL}}\right].
\eeq
It is worth noting that in the case of genuine insulator, $(\gamma_2^A-\gamma_2^B)_{\text{AL}}=-w_{\text{AL}}=0$ since it corresponds to the winding number in the unit cell compatible with the AL.
Once again the chiral polarization field and its relation with the geometric phases allow us to predict the existence of non-trivial zero-energy modes by observing the local discontinuities of the chiral polarization field at any interface.

}

\noindent
{\bf{Chiral polarization in amorphous materials.}}
\label{Appendix:ChiralPolarizationDisorder} We have seen that the chiral polarization does not depend on the specifics of the unit cell: it is an intrinsic property of the material. In fact, as we show below, this framework is far more general and we can define the chiral polarization in amorphous solids.

We start by revisiting the definition of the chiral polarization in a crystal given by eq.~\eqref{eq:PiZak}:

\beq
\bm{\Pi}\equiv 2\sum_{n<0}\expval{\bm x ^A}_{n,\bm R}-\expval{\bm x ^B}_{n,\bm R}.
\eeq

Strictly speaking this polarization is defined at the position $\bm R$. However, the discrete translational invariance of the crystal and by consequence, of the Wannier functions, makes the polarization field homogeneous. We can thus we drop the $\bm R$ indices.

The definition of the Wannier function as the inverse Fourier transform of the Bloch eigenstate cannot be used when dealing with a disordered configuration. Instead, we work with a another set of fully localized functions: the eigenstates of the projected position operator onto the occupied bands~\cite{Neupert:2018}. The projected position operator is given by $P \widehat{X} P$, where
\beq
P = \sum_{E<0} \ket{\Psi_{E}(\bm r)}\bra{\Psi_{E}(\bm r)},
\eeq
is the projector onto the occupied energy states (not to be confused with the projectors $\amsmathbb P^a$), and the $\ket{\Psi_{E}}$ are the eigenstates of the real space hamiltonian $\mathcal H$.
Let us denote the $m^{\rm th}$ eigenstate of the projected position operator as $\widetilde W_m$ (notice that there are as many eigenstates as occupied energy states of the Hamiltonian).
This is a localized function around the center given by $\bm x_m = \expval{\widehat X}{\widetilde W_m}$, similarly to the Wannier centers. Moreover, using each localized function, we can compute the difference of the weighted positions on both sublattices, in other words, the local chiral polarization:
\beq
\bm{\Pi}(\bm x_m) =2\expval{\amsmathbb C \widehat X}{\widetilde W_m}
\label{eq.chiralPolGeneral}
\eeq
In a periodic frame, the eigenstates of the projected position operator reduce to a linear combination of the Wannier functions $W_n$: $\ket{\widetilde W_m}=\sum_n V_{mn} \ket{W_n}$, with $n<0$, indicating the occupied energy bands, $V$ a unitary matrix in the energy space, and $V_{mn}$ a diagonal matrix in the position space. We can then rewrite the chiral polarization in eq.~\eqref{eq.chiralPolGeneral} as

\begin{align}
\bm{\Pi}(\bm x_m)&=2 \sum _{n,l}\bra{W_n}{V^{\dag}_{mn}\amsmathbb C V_{ml}}\ket{W_l}\nonumber\\
&=2\sum _{n<0} \expval{\amsmathbb C\widehat X}{W_n},
\end{align}
where in the last line we used the fact that the $V_{ml}$ commutes with $\amsmathbb C \widehat X$ and the unitarity of $V$. As a result, we recover the first expression defined in crystals using the Bloch formalism as given by Eq.~\eqref{eq:PiZak}.

\noindent
\david{}{
{\bf Chiral polarization, mean chiral displacement and time evolution of Wannier states.}
In Ref.~\cite{maffei2018topological}, the mean chiral displacement under Hamiltonian dynamics was introduced as a measure of the Zak phase of periodic Hamiltonians in $d=1$. 
This quantity characterizes a representation of a Hamiltonian associated to a given unit cell definition, and corresponds to the long-time displacement of an initially fully localized state,
measured  in units of the unit-cell size. As a consequence, several choices of unit cells were necessary to fully characterize the dynamics of a given (meta)material~\cite{cardano2017detection}. 
Although seemingly similar in its formal definition, the chiral polarization which we extensively use in this  article is an intrinsic (meta)material property, defined in real space, and which  does not rely on any underlying frame periodicity, Eq.~\eqref{eq:defPi}. 
In the specific case of periodic frames $\Pi$ crucially resolves the chiral imbalance of wave packets with a sub-unit-cell resolution.

In this method section, we show how $\Pi$ relates to the dynamics of a maximaly localized Wannier state spreading in the bulk of a chiral crystal. To do so 
we consider the time evolution of a wave function $\ket{\psi_n(t) } =  U(t) \ket{W_{n,\bm R}} $ starting from a
of a Wannier state in band $n$, initially localized at ${\bm R}$, 
with an evolution operator $U(t) = \exp (- i H t)$.
Using the notations introduced in Eq.~\eqref{eq:defWannier}, the position at time $t$ is given by
\begin{align}
&\bra{\bm x}\widehat X  \ket{\psi_n(t)} 
\nonumber \\
&=\int_{\bm k}\!\bm x \,e^{i\bm k \cdot(\bm x-\bm R)}
e^{-i E_n({\bm k})t}
{\varphi_{n,\bm k}}(\bm x)
\\
&=\int_{\bm k}\!e^{-i\bm k\cdot \bm R}\left[e^{i\bm k\cdot \bm x}\left(\bm R+\bm v_{n}(\bm k) t + 
i\partial_{\bm k}\right)\right]{\varphi_{n,\bm k}}(\bm x),
\label{eq:xW}
\end{align}
where 
$\bm v_{n}(\bm k) = \partial_{\bm k} E_n({\bm k})$ is the group velocity in the energy band $n$. 
We can also generalize Eq.~\eqref{eq.chiralX2} to  
 define the instantaneous average positions  restricted to the  $a=A,B$ sublattices which read 
\begin{align}
\expval{\bm x ^a (t) }_{n,\bm R}\equiv 
&\expval{\amsmathbb{P}^a \hat X\amsmathbb{P}^a}{\psi_n(t)} \\
=&\oint_{\bm k}\! \frac12 ({\bm R} + {\bm v}_n(\bm k)t )
+\frac{1}{\Omega}\bm{\Gamma}^a_{\text{Zak}}(n)
\\
=& \frac12 {\bm R} 
+\frac{1}{\Omega}\bm{\Gamma}^a_{\text{Zak}}(n)
\\
= & \expval{\bm x ^a (t=0) }_{n,\bm R}. 
\label{eq.chiralX7}
\end{align}
This result indicates that the chiral polarization of each Wannier state is a stationary quantity although they all evolve in time . 
When summed over (half of) the spectrum, we recover the static definition of the chiral polarization
\begin{align}
\bm{\Pi}(t)
&= \textrm{Tr} (U^{-1}(t) \amsmathbb C \hat X U(t) ) \label{eq:PiZak27} \\
&= \sum_n \expval{\amsmathbb{C}  \hat X }{\psi_n(t)}
\\
&= 2 \sum_{n<0}\expval{\bm x ^A (t)}_{n,\bm R}-\expval{\bm x ^B (t)}_{n,\bm R}\nonumber\\
&= 
\frac{2}{\Omega}\sum_{n<0}\bm{\Gamma}^A_{\text{Zak}}(n)-\bm{\Gamma}^B_{\text{Zak}}(n).
\label{eq:PiZak2}
\end{align}
We note that the trace operation in Eq.~\eqref{eq:PiZak27} can be evaluated using any basis of the Hilbert space, such as the ensemble of states fully localized on the $A$ and $B$ sites.
}

\bibliography{articleArxiv}

\begin{thebibliography}{68}
\bibitem [{SSH()}]{SSHWANNIER}%
  \BibitemOpen
  \href@noop {} {\enquote {\bibinfo {title} {For illustrative purposes we used
  a gauge transformation $e^{i\alpha (k)}$, $\alpha(ka)=0.3 \cos(ka)$ to best
  demonstrate the asymmetric spreading of the localized states.}}\
  }\BibitemShut {NoStop}%
\end{thebibliography}

\end{document}